\documentclass[aps,prb,preprintnumbers,english,twocolumn,showpacs]{revtex4}
\usepackage[T1]{fontenc}
\usepackage[latin1]{inputenc}
\usepackage{amsmath}
\usepackage{amssymb}

\providecommand{\LyX}{L\kern-.1667em\lower.25em\hbox{Y}\kern-.125emX\@}

\usepackage{graphicx}
\usepackage{amscd}
\usepackage{bm}

\newcommand{\ud}{\mathrm{d}}
\usepackage{babel}
\begin{document}

\title{Effects of Disorder and Interactions in the Quantum Hall Ferromagnet}
\newcommand{\be}{\begin{equation}}
\newcommand{\ee}{\end{equation}}
\newcommand{\bd}{\begin{displaymath}}
\newcommand{\ed}{\end{displaymath}}

\author{D.\ Makogon, A.\ Alamir, and C.\ Morais Smith}

\affiliation{Institute for Theoretical Physics, University of
Utrecht, Leuvenlaan 4, 3584 CE Utrecht, The Netherlands.}

\date{\today{}}

\begin{abstract}
This work treats the effects of disorder and interactions in a
quantum Hall ferromagnet, which is realized in a two-dimensional
electron gas (2DEG) in a perpendicular magnetic field at Landau
level filling factor $\nu = 1$. We study the problem by projecting
the original fermionic Hamiltonian into magnon states, which
behave as bosons in the vicinity of the ferromagnetic ground
state. The approach permits the reformulation of a strongly
interacting model into a non-interacting one. The latter is a
non-perturbative scheme that consists in treating the two-particle
neutral excitations of the electron system as a bosonic
single-particle. Indeed, the employment of bosonization
facilitates the inclusion of disorder in the study of the system.
It has been shown previously that disorder may drive a quantum
phase transition in the Hall ferromagnet. However, such studies
have been either carried out in the framework of nonlinear sigma
model, as an effective low-energy theory, or included the
long-range Coulomb interaction in a quantum description only up to
the Hartree-Fock level. Here, we establish the occurrence of a
disorder-driven quantum phase transition from a ferromagnetic 2DEG
to a spin glass phase by taking into account interactions between
electrons up to the random phase approximation level in a fully
quantum description.
\end{abstract}


\maketitle
\section{Introduction}

The simultaneous treatment of disorder and interactions in
strongly correlated electron systems has always formed a knotty
challenge; this is because of the dearth of manageable analytical
techniques that can deal with disorder and interactions at the
same time.\cite{green,macdgirvin,chalkerlett} The strongly
correlated system of interest in this work is the two-dimensional
electron gas (2DEG) in a perpendicular magnetic field at Landau
level filling factor $\nu = 1$, whose ground state is commonly
known as the {\it quantum Hall ferromagnet}.

The quantum Hall ferromagnet is the spin-polarized ground state of
the 2DEG at $\nu = 1$ in which all electrons completely fill the
lowest Landau level with spin up polarization. Such configuration
minimizes the Coulomb energy for fermionic systems. In general, it
is a competition between kinetic and Coulomb energies, which
determines the ground state. In the case of the quantum Hall
ferromagnet having $\nu = 1$ the kinetic energy is frozen and does
not change with spin flip, thus, the ground state is
ferromagnetic, even with zero Zeeman splitting. Typically, the
Zeeman splitting in the GaAs heterojunctions turns out to be
roughly 70 times smaller than the spacing between Landau levels
and an order of magnitude smaller than the Coulomb energy per
particle.

The neutral elementary excitations are spin wave excitations, also
called magnons. The spin waves can be described by the action of
the spin lowering operator $S^-_{\bf q}$, projected to the lowest
Landau level, on the ferromagnetic ground state. It turns out,
that the projected operator creates an exact excited eigenstate of
the Hamiltonian. In the regime of low momenta, the magnon's
dispersion is quadratic and the coefficient of the quadratic term
represents a phenomenological constant known as the \textit{spin
stiffness}. The spin stiffness provides a measure of the
free-energy increment associated with twisting the direction of
the spins. A significant spin stiffness indicates that the system
lies in the ferromagnetic phase, while a paramagnetic state
corresponds to a vanishing spin stiffness. The spin wave
dispersion at very large momenta saturates at a constant value
given by the sum of the Coulomb and Zeeman energies. Thus, at
large momenta, the value corresponds to the energy of separate quasiparticle and quasihole excitations. 

One approach that has successfully dealt with strongly correlated
electron systems is the so-called \textit{bosonization} procedure.
Bosonization is a non-perturbative approximation scheme that
essentially treats the electron-hole excitation, known as
\textit{exciton}, as a bosonic single-particle; consequently, a
fermionic Hamiltonian can be recast into a bosonic one. In 1950,
Tomonaga revealed, in a ground-breaking paper,\cite{tomonaga} that
the application of the bosonization formalism to a one-dimensional
electron gas (1DEG) yielded an exactly-solvable Hamiltonian. The
reason is that the electron and the hole propagate with nearly the
same group velocity in the low-energy region. However, that is not
the case in two dimensions. At a given momentum $k$, the
particle-hole pair excitation holds a continuous range of
energies. Therefore, it is less straightforward to construct a
coherently propagating bosonic entity in two dimensions.

The first attempt to extend the bosonization procedure for higher
dimensions was done by Luther \cite{intro2} and then revised by
Haldane.\cite{intro3} Castro Neto and Fradkin,\cite{intro4} as
well as Houghton and Marston,\cite{intro5,intro6} developed a
bosonization technique for a Fermi liquid in any number of
dimensions. As regards the interacting 2DEG subject to an external
perpendicular magnetic field, Westfahl Jr. \textit{et
al.}\cite{D2} constructed a formalism that treated the elementary
neutral excitations of the system, the \textit{magnons}, in a
bosonic framework such that the fermionic Hamiltonian of the
system was transmuted into a quadratic bosonic Hamiltonian. The
drawback is that this method is valid in the limit of weak
magnetic fields, which amounts to large Landau level filling
factors $\nu$.

Doretto \textit{et al.}\cite{doretto} extended the methodology of
Westfahl Jr. \textit{et al.}\cite{D2} to the case of the 2DEG at
$\nu = 1$ (i.e. for a very strong magnetic field). Given that the
system is now restricted to one Landau level, the task greatly
simplifies, since the Landau level quantum degree of freedom can
then be disregarded. Projecting the original fermionic interacting
Hamiltonian of the system into the lowest Landau level, which is
completely filled ($\nu = 1$), allows one to expand it in magnon
states.\cite{Lazarides} It then turns out remarkably that the
dispersion relation of the free magnons coincides with the result
derived by Kallin and Halperin\cite{kallin} within the fermionic
description at the {\it random phase approximation} (RPA) level
and the quartic interacting part of the magnon Hamiltonian might
be related to the skyrmion-antiskyrmion neutral excitations of the
Hall ferromagnet.\cite{doretto} Moreover, in the vicinity of the
ground state, without magnon-magnon interactions, magnons behave
like bosons. This allows to treat magnons approximately as bosons
in the so-called single-mode approximation. \cite{Girvin}

Here, we intend to calculate a quantum phase transition in the quantum Hall
ferromagnet driven by disorder, accounting for the Coulomb interactions between
electrons. We will use the bosonization technique allied to the usual self-consistent
Born-approximation for the disorder averaging procedure.

Before presenting the results obtained in this paper, it is worth
getting acquainted with the current status of research related to
the field. To begin with, Green \cite{green} propounded that the
vanishing of the renormalized spin stiffness at a threshold value
of the disorder strength signifies the occurrence of a
depolarization transition from the ferromagnetic phase to a
paramagnetic one. His finding is based upon a previous result
established by Fogler and Shklovskii, \cite{fogler} who proffered
the same idea in the case of higher Landau levels. Green
established this proposition in the framework of non-linear sigma
model, used as an effective low-energy theory in the regime of
weak disorder.
The other quantity that Green computed is the disorder contribution
to the optical conductivity, which he found to be unmeasurably
small. Finally, Green established that the quantization of the Hall
conductivity is not affected by the presence of weak disorder in the
system.\cite{green}

Another work was carried out by Sinova, MacDonald and
Girvin,\cite{macdgirvin} who established the occurrence of a phase
transition from the paramagnetic state to the partially-polarized
ferromagnetic one and then finally to the fully-polarized
ferromagnetic one as the interaction strength increases relative to
the disorder strength. They determined this result by computing the
average value of the spin polarization as a function of the
interaction strength relative to the disorder strength. Sinova
\textit{et al.}\cite{macdgirvin} did consider Coulomb interactions
within the framework of the Hartree-Fock approximation. Moreover,
the transition from the paramagnetic phase to the ferromagnetic one
was found to take place when the Coulomb energy scale is about twice
as large as the Landau-level-broadening disorder energy scale. As a
final point, the authors inferred that no phase transition can take
place in the strong disorder limit.

The last germane paper was published by Rapsch, Lee and
Chalker.\cite{chalkerlett} They established the occurrence of a
phase transition from the ferromagnetic state to the so-called
spin glass phase. This result was obtained by calculating the
magnetization, the magnetic susceptibility and the spin stiffness
as functions of the disorder strength. They assumed the disorder
potential to be Gaussian distributed and described the system in
terms of a {\it semiclassical} spin model. In their model, they
took into account Coulomb interactions within the Hartree-Fock
approximation but modelled them as being short-ranged. Like
Green,\cite{green} Rapsch \textit{et al.}\cite{chalkerlett}
computed the disorder contribution to the optical conductivity and
found as well that it is undetectable. Finally, they calculated
the dielectric susceptibility of both the partially-polarized
ferromagnetic phase and the spin glass one and they concluded that
both regimes display an insulating behavior at low momenta and a
metallic behavior at large momenta.

Let us now put our work in perspective. Our objective is to
establish the behavior of the renormalized spin stiffness as a
function of the disorder strength in order to ascertain a
potential quantum phase transition driven by disorder to a
non-ferromagnetic state. Indeed, if the spin stiffness vanishes
for a critical value of the disorder strength, then this signals
an instability in the ferromagnetic phase.\cite{green} On the
other hand, the appearance of an imaginary component of the spin
stiffness, which might be interpreted as a spin wave
damping,\cite{Avgin, Shender} at a certain disorder strength,
might indicate the appearance of localized spin waves and a
spin-glass phase transition. Another important characteristic is
the Pauli susceptibility, which diverges at the point of the phase
transition from a non-ferromagnetic to a ferromagnetic state,
indicating spontaneous magnetization. We consider a fully {\it
quantum} model, include a short-range weak disorder potential up
to the $2^{\text{nd}}$ order Born approximation and treat the true
{\it long-range Coulomb} interactions up to the RPA level.

The method that we employ consists of five steps. First of all, a
bosonized expression of the total Hamiltonian, which includes a
contribution from disorder, is sought for. The dispersion relation
of the free bosons corresponds to the one computed by MacDonald
\textit{et al.} \cite{MacDonald} and more explicitly by Doretto
\textit{et al.},\cite{doretto} which entails interactions between
electrons up to the RPA level. The second step consists in
obtaining the full Green's function, and precisely its disorder
self-average. 
In our case, because the impurities are randomly distributed
throughout the system, the disorder self-average can also be taken
by averaging over the impurity positions. The third stage is then
to determine the self-energy of that disorder self-averaged
Green's function through the use of the Dyson's equation. The
self-energy is determined in the low-impurity density and weak
disorder scattering approximations. As a result, the self-energy
corresponds to a single diagram with one propagator line and two
disorder potential lines. The propagator line is evaluated within
two further possible approximations: the bare approximation, which
consists in using the bare bosonic propagator, and the
self-consistent approximation, which uses instead the full
disorder self-averaged Green's function. One must bear in mind
that both propagators take into account interactions between
electrons up to the RPA level. Furthermore, the bare approximation
is first taken in the long wavelength limit, which keeps the
lowest order terms in momenta, and then in the general case, where
all the momenta terms are taken into account. The fourth step
consists in obtaining the renormalized dispersion in these
approximations: bare and self-consistent approximations. The final
stage is then to determine the spin stiffness in the
approximations by taking the coefficient of the quadratic term in
the renormalized dispersion. It is found that a naive
extrapolation of the bare approximation to the regime of finite
disorder strength predicts vanishing of the renormalized spin
stiffness at a certain disorder strength $u_p$, indicating a
paramagnetic phase transition. A more realistic self-consistent
approximation, however, predicts even faster decrease of the
renormalized spin stiffness with growing disorder strength up to a
certain critical value $u_c$ of the disorder. At this point, the
renormalized spin stiffness drastically changes its behavior: it
becomes nonanalytic, acquires an imaginary part, and the real part
saturates at a certain positive value without reaching zero. Such
nonanalytic behavior cannot be accessed by any finite number of
perturbative corrections. In addition, our calculations show a
strong indication that the Pauli susceptibility also diverges at
the same critical point $u_c$, suggesting a phase transition,
presumably to a spin glass phase.

The outline of this paper is the following: in Section II we
present the model and in Section III we derive the expression for
the self-energy. Then, we first solve the problem using the bare
Green's function in Section IV. We present our numerical and
analytical results for the self-consistent solution of the Dyson
equations in Section V and draw our conclusions in Section VI.

\section{The Model}

The 2DEG in the presence of both a perpendicular magnetic field
($\textbf{B} = B \hat{z}$) at $\nu = 1$ and disorder is described
by the fermionic Hamiltonian ${\cal H}={\cal H}_0+{\cal H}_{\rm
imp}$, with
\begin{equation}
\begin{split}
{\cal H}_0 & = \frac{1}{2 m^*} \int \ud \textbf{r} \Psi^{\dagger}
(\textbf{r})
\left( - i \hbar \nabla + e \textbf{A} (\textbf{r}) \right)^2 \Psi (\textbf{r}) \\
& - \frac{1}{2} g^* \mu_B B \sum_{\sigma} \int \ud \textbf{r} \sigma \Psi^{\dagger}
(\textbf{r}) \Psi (\textbf{r}) \\
&+ \frac{1}{2} \sum_{\sigma, \sigma '} \int \ud \textbf{r} \ud
\textbf{r}' \ \Psi^{\dagger}_{\sigma} ( \textbf{r} )
\Psi^{\dagger}_{\sigma '} ( \textbf{r} ' ) V ( |\textbf{r} -
\textbf{r} ' |) \Psi_{\sigma ' } ( \textbf{r} ' ) \Psi_{\sigma} (
\textbf{r} ) \nonumber
\end{split}
\end{equation}
and
\begin{equation}
{\cal H}_{\rm imp} = \int \ud \textbf{r} \sum_{i=1}^{N_{\rm imp}}
U (\textbf{r} - \textbf{X}_i) \Psi^{\dagger} (\textbf{r}) \Psi
(\textbf{r}) . \nonumber
\end{equation}
Here, $\Psi^{\dagger} (\textbf{r})$ and $\Psi (\textbf{r})$ are,
respectively, the fermionic creation and annihilation operators in
coordinate space, $m^*$ denotes the effective mass of the
electron, ${\bf A}$ is the vector potential, $g^*$ stands for the
effective Land\'{e} $g$-factor and $\mu_B$  is the Bohr magneton.
In addition, $V (|\textbf{r}|) = e^2 / ( \epsilon |\textbf{r}| )$
denotes the Coulomb potential, with $\epsilon$ being the
dielectric constant of the host semiconductor, and $U$ stands for
the impurity potential, with $X_i$ being the random position of an
impurity.

The first step consists in obtaining a $2^{\text{nd}}$ quantized
version of the magnon Hamiltonian of the system. In our model we
consider only single magnon processes, which allow us to use a
bosonic description. It was shown in Ref.\ [\onlinecite{doretto}]
that the bosonized Hamiltonian of the system in the absence of
disorder is (neglecting a constant term)
\begin{equation}
{\cal H}_0 = \sum_{\textbf{q}} \omega_{\textbf{q}} b^{\dagger}_{\textbf{q}}
b_{\textbf{q}},
\end{equation}
where $b^{\dagger}_{\textbf{q}}$ and $b_{\textbf{q}}$ are,
respectively, the bosonic creation and annihilation operators in
$\textbf{q}$ space and the bosonic dispersion relation is given by
\begin{equation}\label{berbatovatmantud3}
\omega_{\textbf{q}} = g + \epsilon_B \left[ 1- e^{- |\ell \textbf{q}|^2 / 4}
I_0 \left( \frac{|\ell \textbf{q}|^2}{4} \right) \right].
\end{equation}
Here, $\epsilon_B = \sqrt{\pi / 2} (e^2 / \epsilon \ell)$ stands
for the Coulomb energy scale ($\ell$ being the magnetic length),
$I_0$ denotes the modified Bessel function of the first kind, and
$g = g^* \mu_B B$. It must be stressed that although the
interaction between magnons is omitted from the discussion, the
Coulomb interaction between electrons up to RPA level is taken
into account by the bosonic dispersion relation
$\omega_{\textbf{q}}$.\cite{doretto,kallin}


We now focus on the impurity part of the Hamiltonian. We begin with the fermionic expression
of the $2^{\text{nd}}$ quantized impurity Hamiltonian,
\begin{equation}\label{arshavin2}
{\cal H}_{\rm imp} = \sum_{ \textbf{q}} U (\textbf{q}) g_{\textbf{q}} \sum_{\textbf{p}}
a^{\dagger}_{\textbf{p} + \textbf{q}} a_{\textbf{p}}.
\end{equation}
Here, $g_{\textbf{q}}$ denotes the Fourier transformed density function
$\sum_{j=1}^{N_{\rm imp}} \delta (\textbf{x} - \textbf{X}_j)$ for the impurities and
$a^{\dagger}_{\textbf{q}}$ and $a_{\textbf{q}}$ are, respectively, the fermionic
creation and annihilation operators in $\textbf{q}$ space. In order to obtain the
bosonic form of the above, the Fourier-transformed electronic density operator must
be used. It is given by
\begin{equation}\label{arshavin1}
\rho (\textbf{q}) = \int \ud \textbf{r} \ e^{- i \textbf{q} \cdot \textbf{r}}
\Psi^{\dagger} (\textbf{r}) \Psi (\textbf{r}).
\end{equation}
The electronic field operators are related to the single-electron operators by
\begin{equation} \nonumber
\Psi (\textbf{r}) = \sum_{\textbf{p}} \frac{e^{-i \textbf{p} \cdot
\textbf{r}}}{\sqrt{A}} a_{\textbf{p}} \ \ \text{and} \ \
\Psi^{\dagger} (\textbf{r}) = \sum_{\textbf{p}}\frac{e^{i
\textbf{p} \cdot \textbf{r}}}{\sqrt{A}} a^{\dagger}_{\textbf{p}},
\end{equation}
where $A$ is the area of the system. Substituting the above back
into Eq.\;(\ref{arshavin1}) gives
\begin{equation}
\rho (\textbf{q}) = \int \ud \textbf{r} \ e^{- i \textbf{q} \cdot \textbf{r}}
 \sum_{\textbf{p}, \textbf{p}'} \frac{e^{i (\textbf{p}'- \textbf{p}) \cdot
\textbf{r}} }{A} a^{\dagger}_{\textbf{p}'} a_{\textbf{p}} = \sum_{\textbf{p}}
a^{\dagger}_{\textbf{p} + \textbf{q}} a_{\textbf{p}}.
\label{eq7}
\end{equation}
Then, substituting Eq.\ (\ref{eq7}) back into Eq.\;(\ref{arshavin2}) yields
\begin{equation}
{\cal H}_{\rm imp} = \sum_{ \textbf{q}} U (\textbf{q}) g_{\textbf{q}} \rho (\textbf{q}).
\end{equation}
The bosonized version of the electron density operator reads \cite{doretto}
\begin{equation}
\rho (\textbf{q}) = \delta_{\textbf{q}, 0} N_{\phi} + 2 i e^{-|\ell \textbf{q}|^2 /4}
\sum_{\textbf{p}} \sin \left( \frac{\textbf{q} \wedge \textbf{p}}{2} \right)
b^{\dagger}_{\textbf{q}+ \textbf{p}} b_{\textbf{p}},
\end{equation}
where $N_{\phi}= A/(2 \pi l^2) $ is the Landau level degeneracy
and $\textbf{q} \wedge \textbf{p} = \ell^2 \hat{z} \cdot
(\textbf{q} \times \textbf{p})$. The disorder Hamiltonian then
becomes
\begin{eqnarray} \nonumber
{\cal H}_{\rm imp} & = \sum_{ \textbf{q}} U (\textbf{q}) g_{\textbf{q}} \left[
\delta_{\textbf{q}, 0} N_{\phi} + 2 i e^{-|\ell \textbf{q}|^2 /4}\right.  \\
& \left.  \times \sum_{\textbf{p}}
\sin \left( \frac{\textbf{q} \wedge \textbf{p}}{2} \right) b^{\dagger}_{\textbf{q}+
\textbf{p}} b_{\textbf{p}} \right].
\end{eqnarray}
The constant term $\delta_{\textbf{q}, 0} N_{\phi}$ is now omitted since the
quantity of interest is the Green's function.

The bosonized impurity Hamiltonian is then finally written as
\begin{equation}
{\cal H}_{\rm imp} = \sum_{ \textbf{q} , \textbf{p}} U (\textbf{q}) g_{\textbf{q}}
f (\textbf{q}, \textbf{p}) b^{\dagger}_{\textbf{q}+ \textbf{p}} b_{\textbf{p}},
\end{equation}
where
\begin{equation}
f (\textbf{q}, \textbf{p}) = 2 i e^{-|\ell \textbf{q}|^2 /4} \sin \left(
\frac{\textbf{q} \wedge \textbf{p}}{2} \right).
\end{equation}

Labelling
\begin{equation}\label{qzdr1}
U (\textbf{q}) f (\textbf{q}, \textbf{p})  = U^{e} (\textbf{q},
\textbf{p}),
\end{equation}
the full bosonized Hamiltonian of the quantum Hall ferromagnet in the presence of
impurities is then expressed as
\begin{equation}\label{arda1}
{\cal H} = \sum_{\textbf{q}} \omega_{\textbf{q}}
b^{\dagger}_{\textbf{q}} b_{\textbf{q}} + \ \sum_{ \textbf{q} ,
\textbf{p}} U^{e} (\textbf{q}, \textbf{p}) g_{\textbf{q}}
b^{\dagger}_{\textbf{q}+ \textbf{p}} b_{\textbf{p}}.
\end{equation}

Let us now say a few words on the dimensions of the disorder potential. There are two
sources of disorder present in the system: impurities positioned at a certain
distance away from the 2DEG and impurities present in the 2DEG. In the case of GaAs
heterostructures, \cite{green,murthy} most of the disorder potential is spawned by
the Coulomb interaction between the electrons and the impurities located away from
the 2DEG. These impurities correspond to ionized donor atoms situated in the $n$-type
region, which itself is detached from the 2DEG by an insulating layer of thickness
$d \sim 1000 \text{\AA{}} \gg \ell$. In
the present calculations, the disorder potential will be taken as an effective
two-dimensional potential.

Having obtained the bosonized Hamiltonian in the presence of impurities, one is now
able to determine the expression for the self-energy.

\section{Derivation of the self-energy}

In the same spirit as Ref [\onlinecite{doniach}], one first looks
for the Green's function, 
\begin{equation}\label{Greendoniach}
G (\textbf{p}', \textbf{p} ; t) = - i \langle 0 | T [
b_{\textbf{p}} (t) b^{\dagger}_{\textbf{p}'} (0) ] | 0 \rangle.
\end{equation}
Here, $| 0 \rangle$ stands for the bosonic vacuum state, which is
none other than the quantum Hall ferromagnet: i.e. $| 0 \rangle
\equiv | QHF \rangle = \prod_{m=0}^{N_{\phi}-1} c^{\dagger}_{m,
\uparrow} | 0 \rangle_F$. Thus, one has
\begin{equation}\label{arda2}
i \frac{\partial}{\partial t} G (\textbf{p}', \textbf{p} ; t) =
\delta (t) \delta_{\textbf{p} , \textbf{p}'} - i \langle 0 | T
\left[ [b_{\textbf{p}} (t) ,{\cal H}]b^{\dagger}_{\textbf{p}'} (0)
\right] | 0 \rangle ,
\end{equation}
where $T$ is the time ordering operator. Now, using
Eq.\;(\ref{arda1}), one easily finds that
\begin{equation}
[b_{\textbf{p}} (t), {\cal H}] = \omega_{\textbf{p}} b_{\textbf{p}}
(t) + \sum_{\textbf{q}} U^{e} (\textbf{q}, \textbf{p} - \textbf{q})
g_{\textbf{q}} b_{\textbf{p} - \textbf{q}} (t),
\end{equation}
such that one obtains for the second term in Eq.\;(\ref{arda2})
\begin{eqnarray} \nonumber
& - i \langle 0 | T \left[ [b_{\textbf{p}} (t) ,{\cal H}]b^{\dagger}_{\textbf{p}'} (0)
\right] | 0 \rangle = \omega_{\textbf{p}} G (\textbf{p}', \textbf{p};t) \\
& + \sum_{\textbf{q}} U^{e}(\textbf{q}, \textbf{p} - \textbf{q})
g_{\textbf{q}} G (\textbf{p}' , \textbf{p} - \textbf{q};t).
\end{eqnarray}
Hence, the equation of motion of $ G (\textbf{p}', \textbf{p} ;
t)$ is written as
\begin{eqnarray} \nonumber &&
\left( i \frac{\partial}{\partial t} - \omega_{\textbf{p}} \right)
G (\textbf{p}',
\textbf{p} ; t) = \delta_{\textbf{p}, \textbf{p}'} \delta (t) \\
&+& \sum_{\textbf{q}} U^{e} (\textbf{q}, \textbf{p} - \textbf{q})
g_{\textbf{q}} G (\textbf{p}' , \textbf{p} -
\textbf{q};t).\label{arda2bis}
\end{eqnarray}

The zero-order approximation to the solution of Eq.\;(\ref{arda2bis}) yields
\begin{equation}\label{arda3}
G^0 (\textbf{p}', \textbf{p} ; t) = \delta_{\textbf{p},
\textbf{p}'} G^0 (\textbf{p}, t),
\end{equation}
where $G^0 (\textbf{p}, t)$ stands for the bare bosonic Green's function.
We now look for the expression for $G^0 (\textbf{p}, t)$.

Firstly, one needs to find the Heisenberg bosonic operator in the
absence of the disorder potential. Starting with $ i \partial_t b_{\textbf{p}} (t) =
[b_{\textbf{p}} (t), {\cal H}_0] = \omega_{\textbf{p}} b_{\textbf{p}} (t) $, one then
obtains $ b_{\textbf{p}} (t) = b_{\textbf{p}} e^{-i \omega_{\textbf{p}} t} $.
Therefore, for the case $t > 0$, the free Green's function is
\begin{eqnarray} \nonumber
G^0 (\textbf{p}, t) &=& -i \langle 0 | b_{\textbf{p}}(t)
b^{\dagger}_{\textbf{p}} | 0 \rangle = -i e^{-i \omega_{\textbf{p}}
t} \langle 0 | b_{\textbf{p}} b^{\dagger}_{\textbf{p}} | 0 \rangle
\\\nonumber &=& - i e^{-i \omega_{\textbf{p}} t},
\end{eqnarray}
whereas for $t < 0$, it turns out to be
\begin{equation}\nonumber
G^0 (\textbf{p}, t) = -i \langle 0 | b_{\textbf{p}}(t)
b^{\dagger}_{\textbf{p}} | 0 \rangle = -i e^{-i \omega_{\textbf{p}}
t} \langle 0 | b^{\dagger}_{\textbf{p}} b_{\textbf{p}} | 0 \rangle =
0.
\end{equation}
This solution is indeed identical to the electronic one.

Now, the cynosure is on the generic solution of the differential equation
(\ref{arda2bis}). By coupling the latter with the boundary equation (\ref{arda3})
yields the integral equation
\begin{equation}\label{arda4}
\begin{split}
G (\textbf{p}', \textbf{p} ; t) &= \delta_{\textbf{p},
\textbf{p}'} G^0 (\textbf{p},t)
+ \int^{\infty}_{- \infty} \ud t' \ G^0 (\textbf{p}, t - t') \\
& \times \sum_{\textbf{q}} U^{e} (\textbf{q}, \textbf{p} -
\textbf{q}) g_{\textbf{q}} G (\textbf{p}', \textbf{p} - \textbf{q};
t).
\end{split}
\end{equation}
By Fourier-transforming the time in Eq.\;\eqref{arda4} to
frequency and shifting $\textbf{q} \rightarrow \textbf{p} -
\textbf{q}$ one finds
\begin{equation}\label{arda5}
\begin{split}
G (\textbf{p}', \textbf{p}; \omega) & = \delta_{\textbf{p},
\textbf{p}'}
G^0 (\textbf{p}, \omega) + G^0 (\textbf{p}, \omega ) \\
& \times \sum_{\textbf{q}} U^{e} ( \textbf{p} - \textbf{q},
\textbf{q}) g_{\textbf{p} - \textbf{q}} G (\textbf{p}', \textbf{q};
\omega).
\end{split}
\end{equation}
Here, the bare Green's function reads
\begin{equation}\label{berbatovatmanutd1}
G_0 (\textbf{p} , \omega) = \frac{1}{\omega - \omega_{\textbf{p}} +
i \eta},
\end{equation}
where $\eta \rightarrow 0^+$ and $\omega_{\textbf{p}}$ is given by
Eq.\;\eqref{berbatovatmantud3}. In the same way as for fermions,
the solution of Eq.\;(\ref{arda5}) is obtained by iteration. One
gets the so-called \textit{Born series}:
\begin{equation}\label{saqibismyfriend1}
G (\textbf{p}', \textbf{p}) = \sum_{n=0}^{\infty} G^{(n)}
(\textbf{p}', \textbf{p}),
\end{equation}
where $G^0 (\textbf{p}', \textbf{p}) = \delta_{\textbf{p},
\textbf{p}'} G^0 (\textbf{p})$ and for $n \geq 1$,
\begin{equation}\nonumber
G^{(n)} (\textbf{p}', \textbf{p}) = G^0 (\textbf{p})
\sum_{\textbf{q}} U^{e} (\textbf{p} - \textbf{q}, \textbf{q})
g_{\textbf{p} - \textbf{q}} G^{(n - 1)} (\textbf{p}' , \textbf{q}).
\end{equation}
Expansion of Eq.\;(\ref{saqibismyfriend1}) then yields
\begin{widetext}
\begin{equation}\nonumber
\begin{split}
G (\textbf{p}', \textbf{p}) &  = \delta_{\textbf{p}, \textbf{p}'}
G^0 (\textbf{p}') + G^0 (\textbf{p}') U^{e} (\textbf{p} -
\textbf{p}', \textbf{p}') g_{\textbf{p} - \textbf{p}'} G^0
(\textbf{p}) + \sum_{\textbf{q}} G^0 (\textbf{p}') U^{e} (\textbf{q}
- \textbf{p}', \textbf{p}') g_{\textbf{q} - \textbf{p}'} G^0
(\textbf{q}) U^{e} (\textbf{p} -
\textbf{q}, \textbf{q}) g_{\textbf{p} - \textbf{q}} G^0 (\textbf{p})  \\
&  + \sum_{\textbf{q}, \textbf{q}'} G^0 (\textbf{p}')
U^{e}(\textbf{q}- \textbf{p}', \textbf{p}') g_{\textbf{q}-
\textbf{p}'} G^0 (\textbf{q}) U^{e} (\textbf{q}' - \textbf{q},
\textbf{q}) g_{\textbf{q}' - \textbf{q}} G^0 (\textbf{q}') U^{e}
(\textbf{p} - \textbf{q}', \textbf{q}') g_{\textbf{p} - \textbf{q}'}
G^0 (\textbf{p})  + \ldots,
\end{split}
\end{equation}
\begin{figure}[t]
\centering
\includegraphics[scale= 0.37]{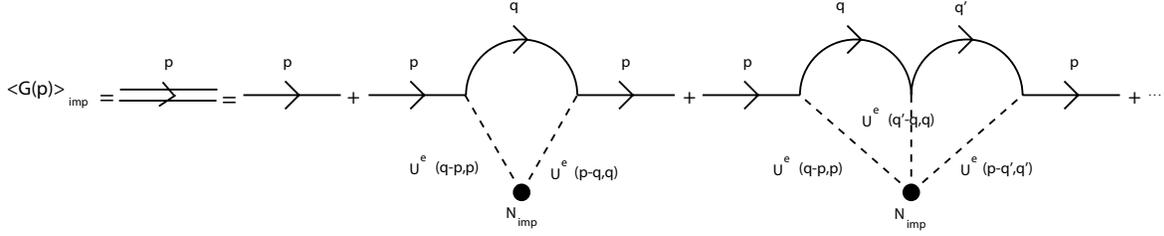}
\caption{Diagrammatic expansion of the disorder averaged Green's function.}
\label{diaag9}
\end{figure}

Due to disorder self-averaging in the limit of very large number of
impurities $N_{\rm imp}\rightarrow \infty$, with constant density
$n_{\rm imp}={\rm const.}$, the full bosonic one-particle Green's
function approaches its average value
\begin{equation}
\langle (G (\textbf{p}', \textbf{p})-\langle G (\textbf{p}',
\textbf{p}) \rangle_{\rm imp})^2\rangle_{\rm imp}\rightarrow 0,
\end{equation}
which is
\begin{equation}\nonumber
\begin{split}
\langle G (\textbf{p}', \textbf{p}) \rangle_{\rm imp} &  =
\delta_{\textbf{p}, \textbf{p}'} G^0 (\textbf{p}') + \langle
g_{\textbf{p} - \textbf{p}'} \rangle_{\rm imp}G^0 (\textbf{p}')
U^{e} (\textbf{p} - \textbf{p}', \textbf{p}') G^0 (\textbf{p})
\\& + \sum_{\textbf{q}}\langle g_{\textbf{q} -
\textbf{p}'}g_{\textbf{p} - \textbf{q}} \rangle_{\rm imp} G^0
(\textbf{p}') U^{e} (\textbf{q} - \textbf{p}', \textbf{p}')
 G^0 (\textbf{q}) U^{e}
(\textbf{p} - \textbf{q}, \textbf{q}) G^0 (\textbf{p})  \\
&  + \sum_{\textbf{q}, \textbf{q}'}\langle g_{\textbf{q}-
\textbf{p}'}g_{\textbf{q}' - \textbf{q}}g_{\textbf{p} -
\textbf{q}'}\rangle_{\rm imp}G^0 (\textbf{p}') U^{e}(\textbf{q}-
\textbf{p}', \textbf{p}')  G^0 (\textbf{q}) U^{e} (\textbf{q}' -
\textbf{q}, \textbf{q})
 G^0 (\textbf{q}') U^{e}
(\textbf{p} - \textbf{q}', \textbf{q}')  G^0 (\textbf{p})  + \ldots
\end{split}\label{arda6}
\end{equation}
In the thermodynamic limit $A\rightarrow \infty$:
\begin{align}
\langle g_{\textbf{q}} \rangle_{\rm imp} &  = N_{\rm imp} \delta_{\textbf{q},0},
\nonumber \\
\langle g_{\textbf{q}} g_{\textbf{p}} \rangle_{\rm imp} & =
N^2_{\rm imp} \delta_{\textbf{p}, \textbf{0}}\delta_{\textbf{q},0} + N_{\rm imp}\delta_{\textbf{q}+\textbf{p},0}, \nonumber \\
\langle g_{\textbf{q}'} g_{\textbf{q}} g_{\textbf{p}} \rangle_{\rm
imp}  &  = N^3_{\rm imp}\delta_{\textbf{q}', 0}
\delta_{\textbf{q}, 0}\delta_{\textbf{p}, 0} + N^2_{\rm imp}
(\delta_{\textbf{p}+\textbf{q}, 0}\delta_{\textbf{q}', 0} +
\delta_{\textbf{q}+ \textbf{q}',0}\delta_{\textbf{p}, 0} +
\delta_{\textbf{p}+ \textbf{q}',0}\delta_{\textbf{q}, 0}) + N_{\rm
imp}\delta_{\textbf{q}'+\textbf{q}+\textbf{p}, 0}.
\label{kasmirisvital}
\end{align}
Moreover, one has
\begin{equation}
U^{e} (0, \textbf{p}) = U (0) f (0, \textbf{p}) = U (0) 2 i e^{-
|\ell (0)|^2 / 4} \sin \left( \frac{0 \wedge \textbf{p} }{2} \right)
= 0.\label{firstorder}
\end{equation}
Substituting Eqs.\;\eqref{kasmirisvital} and Eq.\;(\ref{firstorder})
into the expression for $\langle G (\textbf{p}) \rangle_{\rm imp}$
 shows that the translational invariance is
recovered after the averaging $\langle G (\textbf{p}', \textbf{p})
\rangle_{\rm imp}=\langle G (\textbf{p}) \rangle_{\rm
imp}\delta_{\textbf{p}', \textbf{p}}$, where
\begin{equation}
\begin{split}
\langle G (\textbf{p}) \rangle_{\rm imp} &  = G^0 (\textbf{p}) +
N_{\rm imp} \sum_{\textbf{q}} G^0 (\textbf{p}) U^{e} (\textbf{q} -
\textbf{p}, \textbf{p}) G^0 (\textbf{q}) U^{e} (\textbf{p} -
\textbf{q},
\textbf{q}) G^0 (\textbf{p}) \\
& \ \quad + N_{\rm imp} \sum_{\textbf{q}, \textbf{q}'} G^0
(\textbf{p}) U^{e} (\textbf{q} - \textbf{p}, \textbf{p}) G^0
(\textbf{q}) U^{e} (\textbf{q}' - \textbf{q}, \textbf{q})  G^0
(\textbf{q}') U^{e} (\textbf{p} - \textbf{q}', \textbf{q}') G^0
(\textbf{p}) + \dots.
\end{split}
\end{equation}
Therefore, there is no $1^{st}$ order Born scattering contribution
to the bosonic self-energy. Moreover, it is possible to show that
all odd order contributions to the self-energy vanish (see
Appendix A).

\end{widetext}

This result is expressed diagrammatically in Fig.\;\ref{diaag9}.
It was shown\cite{doniach} that the disorder averaged Green's function can also be
expressed as
\begin{equation}\label{berabatamanutd4}
\langle G (\textbf{p}) \rangle_{\rm imp} = \frac{1}{\omega - \omega_{\textbf{p}} -
\Sigma(\textbf{p}, \omega) }.
\end{equation}

\begin{figure}[t] \centering
\includegraphics[scale= 0.4]{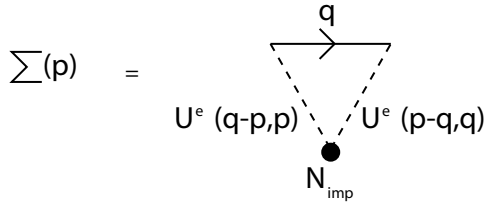}
\caption{Self-energy in the low-density weak scattering
approximation.}
\label{diaag10}
\end{figure}

Hence, the self-energy must now be computed. The low-density weak
scattering approximation will be used throughout the calculations.
Low density means that the number of disorder atoms present in the
system is taken to be much lower than the number of electrons,
while the weak-scattering approximation signifies that the
scattering potential induced by a given impurity atom is weak,
such that only the first and second-order Born scatterings are
accounted for. The problem then reduces to solving the
diagrammatic expression shown in
Fig.\;\ref{diaag10}.

The self-energy can be evaluated in two different manners: 1) the
bare approximation that uses the bare propagator $G^0$ and 2) the
self-consistent approximation that uses the full disorder
self-averaged Green's function $\langle G \rangle _{\rm imp}$.
Therefore, for generality we will use the propagator $G (\textbf{q},
\omega)$, which is going to be specified further for each particular
case. This yields algebraically
\begin{widetext}
\begin{equation}
\begin{split}
\Sigma (\textbf{p}, \omega) &  = N_{\rm imp} \sum_{\textbf{q}} U^{e}
(\textbf{q} - \textbf{p}, \textbf{p}) G (\textbf{q}, \omega) U^{\rm
e} (\textbf{p} - \textbf{q},
\textbf{q} ) \\
& = N_{\rm imp} \sum_{\textbf{q}} U (\textbf{q} - \textbf{p} ) f
(\textbf{q} - \textbf{p}, \textbf{p}) G (\textbf{q}, \omega) U
(\textbf{p} - \textbf{q}) f(\textbf{p} - \textbf{q}, \textbf{q}),
\end{split}
\end{equation}
where Eq.\;(\ref{qzdr1}) was substituted in the second line.

In this work, the impurity potential is assumed to be short-range, i.e.
$U (\textbf{q}) = \text{constant}$. An uniform potential in momentum space is attained
from a delta function interaction in real space, $U (\textbf{r}) = U \delta
(\textbf{r})$ (such that $U (\textbf{q}) = (1/A) \int d \textbf{r} e^{i \textbf{q}
\cdot \textbf{r}} U \delta (\textbf{r}) = U /A$). Thus, this model assumes that the bosons (and therefore the electrons)
collide directly with the impurity 'atoms' that constitute the effective disorder
potential; in reality, most of the impurities are located away from the 2DEG. Thus,
one has
\begin{equation}
\begin{split}
\Sigma (\textbf{p}, \omega) &  = N_{\rm imp} \sum_{\textbf{q}}
\left( \frac{U}{A} \right)^2 \left[ 2 i e^{- |\ell (\textbf{q} -
\textbf{p})|^2 /4} \sin \left( \frac{(\textbf{q} - \textbf{p})
\wedge \textbf{p}}{2} \right) \right] G (\textbf{q}, \omega)
\left[ 2 i e^{- |\ell (\textbf{p} - \textbf{q})|^2 / 4} \sin
\left( \frac{(\textbf{p} -
\textbf{q}) \wedge \textbf{q}}{2} \right) \right] \\
& = 4 N_{\rm imp} \left( \frac{U}{A} \right)^2  \sum_{\textbf{q}}
e^{- |\ell (\textbf{q} - \textbf{p})|^2 / 2} \sin ^2 \left(
\frac{\textbf{q} \wedge \textbf{p}}{2} \right) G (\textbf{q},
\omega ).
\end{split}\label{fghw1}
\end{equation}
One then expands the argument of the exponential:
\begin{equation}
e^{- |\ell (\textbf{q} - \textbf{p})|^2 /2} = e^{- |\ell \textbf{q}|^2 / 2}e^{- |\ell
\textbf{p}|^2 / 2}e^{\ell ^2 \textbf{q} \cdot \textbf{p}} = e^{-(\ell q)^2 /2}
e^{-(\ell p)^2 /2} e^{\ell^2 q p \cos \phi}.
\end{equation}
Here, $\phi$ denotes the angle between vectors $\textbf{q}$ and
$\textbf{p}$. Furthermore, the summation is transmuted into an
integration through the use of the formula,
\begin{equation}
\sum_{\textbf{q}} = \frac{A}{4 \pi^2} \int \ud^2 q = \frac{A}{4 \pi^2} \int^{\infty}_0
\ud q \ q \int^{2 \pi}_0 d \theta.
\end{equation}
The angle $\theta$ is taken arbitrarily on the plane containing
the vector $\textbf{q}$, therefore, one is free to set $\theta =
\phi$. The sine squared term in Eq.\;(\ref{fghw1}) can be
re-written as $\sin^2 ( \textbf{q} \wedge \textbf{p} / 2 ) = [ 1 -
\cos (\textbf{q} \wedge \textbf{p}) ] / 2 = [1 - \cos (\ell^2 q p
\sin \phi) ] / 2$. We also assume rotation invariance of the
Green's function $G (\textbf{q}, \omega )=G (q, \omega )$. Hence,
the self-energy is also rotation invariant and can be expressed as
\begin{equation}\label{fghw2}
\Sigma (p, \omega) = 4 n_{\rm imp} U^2 \int^{\infty}_0 \frac{\ud
q}{2 \pi} \ q e^{-(\ell q)^2 /2} e^{-(\ell p)^2 /2} G (q, \omega )
\int^{2 \pi}_0 \frac{\ud \phi }{2 \pi} \ e^{\ell^2 q p \cos \phi}
\frac{1}{2} [1 - \cos (\ell^2 q p \sin \phi) ],
\end{equation}
where $n_{\rm imp} = N_{\rm imp} / A$ stands for the impurity
density. After a straightforward calculation (see Appendix B), we
find
\begin{equation}\label{fghw3}
\Sigma (p, \omega) = 4 n_{\rm imp} U^2 \int^{\infty}_0 \frac{\ud
q}{2 \pi} \ q e^{-(\ell q)^2 /2} e^{-(\ell p)^2 /2} G (q, \omega )
\frac{1}{2}[ I_0 (\ell^2 q p)-1].
\end{equation}
Rescaling the momenta by $\textbf{q}, \textbf{p} \rightarrow
\textbf{q} / \ell, \textbf{p} / \ell$ simplifies the self-energy
to
\begin{equation}\label{fghw4}
\Sigma (p, \omega) = \frac{u}{4} \epsilon_B^2 e^{-p^2 /
2}\int^{\infty}_0 \ud q \ q e^{-q^2 / 2}[I_0 (q p) -1] G (q,
\omega ),
\end{equation}
where the various pre-factors, including the disorder potential
strength and the impurity density, can be re-grouped into a single
convenient parameter:
\begin{equation}
u = \frac{4n_{\rm imp} U^2}{\pi \ell^2 \epsilon_B^2 },
\label{scottburray1}
\end{equation}
which will be dubbed the \textit{disorder strength}. Thus, $u$ is
a dimensionless parameter that measures the disorder interaction
strength relative to the Coulomb interaction, $u \approx (E_{\rm
dis} / E_{\rm coul})^2$. The above self-energy expression will be
evaluated in two different ways: i) first order corrections in $u$
and ii) self- consistently.

\subsection{Bare  Approximation} 
In the bare approximation the self-energy (\ref{fghw4}) becomes
\begin{equation}\label{fghw1ba}
\Sigma (p, \omega) = \frac{u}{4} \epsilon_B^2 e^{-p^2 /
2}\int^{\infty}_0 \ud q \ q e^{-q^2 / 2}[I_0 (q p) -1] G^0 (q,
\omega ).
\end{equation}
After substituting Eq.\;(\ref{berbatovatmanutd1}) into the above, we
obtain
\begin{equation}\label{fghw2ba}
\Sigma (p, \omega) = \frac{u}{4} \epsilon_B^2 e^{-p^2 /
2}\int^{\infty}_0 \ud q \ q e^{-q^2 / 2}\frac{I_0 (q p)-1 }{\omega
- \omega_{\textbf{q}} + i \eta}.
\end{equation}
Making use of the identity (for $\eta \rightarrow 0^+$),
\begin{equation}
\frac{1}{x + i \eta} = \mathcal{P} \frac{1}{x} - i \pi \delta (x),
\end{equation}
we find the real and imaginary parts of the self-energy
\begin{align}
\text{Re} \Sigma (p, \omega) \  &  = \frac{u}{4} \epsilon_B^2
e^{-p^2 / 2} \mathcal{P} \int^{\infty}_0 \ud q \ q e^{-q^2 / 2}
\frac{ I_0 (q p) -1 }{\omega - \omega_{q} } ,
\label{saaidismybro2} \\
\text{Im} \Sigma (p, \omega) \  &  = - \frac{u}{4} \epsilon_B^2
e^{-p^2 / 2} \int^{\infty}_0 \ud q \ q e^{-q^2 / 2} [I_0 (q p) -1]
\pi \delta (\omega - \omega_{q} ) . \label{saaidismybro2bisbis}
\end{align}

The above equations can be evaluated analytically in the long
wavelength approximation, which is done in Appendix C. Here, one
uses the complete bosonic dispersion relation given by
Eq.\;\eqref{berbatovatmantud3}. As a result, one can only solve
the imaginary self-energy numerically; that task is not performed
here. We concentrate, instead, on the real part.

The renormalized energy of the bosons (including the disorder
contribution) is obtained by looking at the poles of the full
disorder self-averaged Green's function in
Eq.\;\eqref{berabatamanutd4}, $ \omega - \omega_{\textbf{p}} -
\text{Re} \Sigma (\textbf{p}, \omega) = 0 $, such that the
renormalized dispersion relation is determined from
Eq.\;(\ref{saaidismybro2}):
\begin{equation}\label{kirilishere1}
\omega = g + \epsilon_B \left[  1 - e^{- p^2 / 4} I_0 \left(
\frac{p^2}{4} \right) \right]  + \frac{u}{4} \epsilon_B^2 e^{-p^2
/ 2} \mathcal{P} \int^{\infty}_0 \ud q \ q e^{-q^2 / 2} \frac{I_0
(q p) -1}{\omega - \left[ g + \epsilon_B \left( 1 - e^{- q^2 / 4}
I_0 \left( \frac{q^2}{4} \right) \right) \right] }.
\end{equation}
The corresponding plot is illustrated on
Fig.\;\ref{scottmurray3}(a). One can notice that at not too large
momenta (i.e. near $|p \ell| = 1$) there exists already a
substantial difference between the bare (long wavelength) and bare
(full $k$) approximations.
\begin{figure}[t!]
\centering
\includegraphics[scale=0.63]{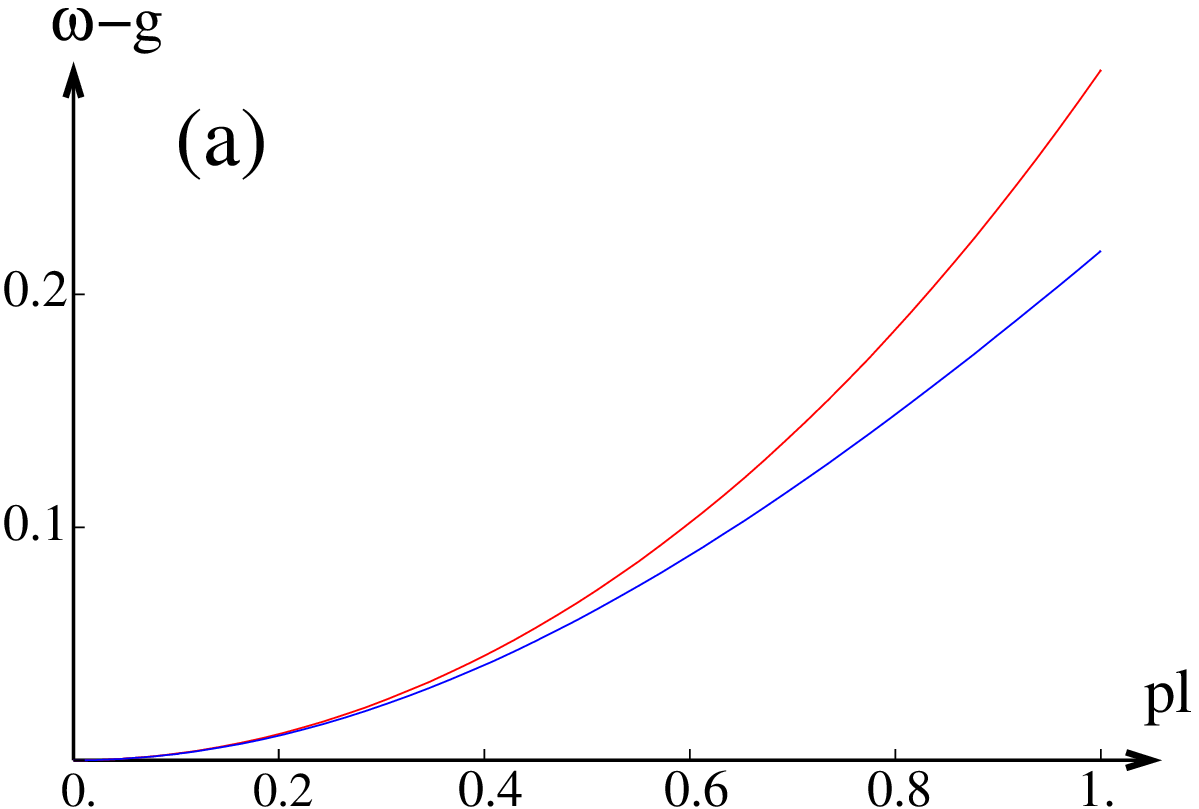}
\includegraphics[scale=0.75]{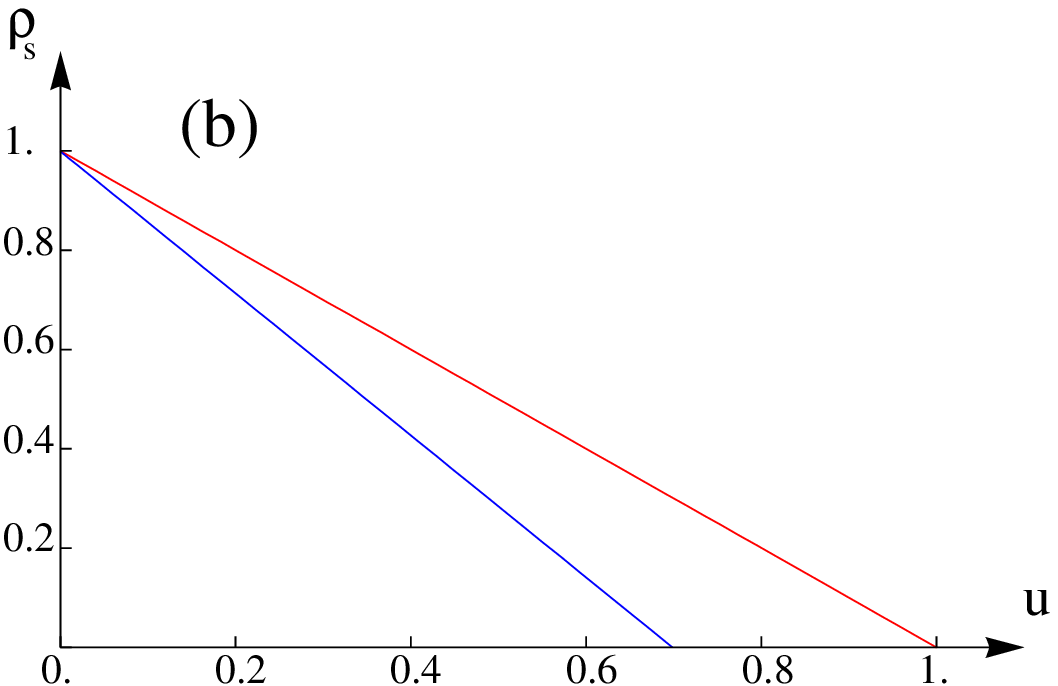}
\caption{(Color online) (a) Renormalized dispersion in the bare
full $k$ approximation (blue), in units of the Coulomb energy $e^2
/ (\epsilon \ell)$, contrasted with the one in the bare long
wavelength approximation (red), both as functions of the momentum
$|p \ell|$ and at $u = 0.1$. (b) Renormalized spin stiffness in
the bare full $k$ approximation (blue) and in the bare long
wavelength approximation (red). Notice that using the bare Green's
function $G_0$ we find a transition from a ferromagnetic to a
paramagnetic phase, whereas using $G$ in the self-consistent
approximation we find a transition into a spin glass phase (see
next section).} \label{scottmurray3}
\end{figure}

Now, the renormalized spin stiffness is sought for. For the sake of convenience,
one begins by introducing the variables $\tilde{\omega}, \tilde{g} = \omega /
\epsilon_B , g / \epsilon_B$ and re-writing Eq.\;(\ref{kirilishere1}) as
\begin{equation}
\tilde{\omega} - \tilde{g} = 1 - e^{- p^2 / 4} I_0
\left(\frac{p^2}{4} \right) + \frac{u}{4} e^{-p^2 / 2} \mathcal{P}
\int^{\infty}_0 \ud q \ q e^{-q^2 / 2} \frac{I_0 (q p)
-1}{(\tilde{\omega} - \tilde{g}) - \left[ 1 - e^{- q^2 / 4} I_0
\left(\frac{q^2}{4} \right) \right] }.
\end{equation}
One then expands the above in powers of p,
\begin{equation}
\begin{split}
\tilde{\omega} - \tilde{g} &  = 1 - \left(1 - \frac{p^2}{4} + \ldots \right)
\left( 1 + \frac{p^4}{64} + \ldots \right) \\
& \ \quad + \frac{u}{4} \left(1 - \frac{p^2}{2} + \dots \right)
\mathcal{P} \int^{\infty}_0 \ud q \ q e^{-q^2 / 2} \frac{  \left(
1 + \frac{q^2 p^2}{4} + \ldots -1 \right)}{(\tilde{\omega} -
\tilde{g}) - \left[ 1 - e^{- q^2 / 4} I_0 \left(\frac{q^2}{4}
\right) \right] },
\end{split}
\end{equation}
and one takes only the $p^2$ terms,
\begin{equation}
\begin{split}
\tilde{\omega} - \tilde{g} &  = \frac{p^2}{4} - \frac{u}{4}
\frac{p^2}{4} \int^{\infty}_0 \ud q \ \frac{q^3 e^{-q^2 /
2}}{\left[1 - e^{-q^2 /4} I_0
\left( \frac{q^2}{4} \right) \right]}  + \ldots \\
& = \frac{p^2}{4} \left[1 - \frac{u}{4} (5.72) \right] + \dots
\end{split}
\end{equation}
The renormalized spin stiffness then reads
\begin{equation}\label{mainbare}
\rho_s = \frac{\epsilon_B}{4} (1 - 1.43 u).
\end{equation}
Eq.\;(\ref{mainbare}) is the main result of this section. The
above expression was derived in the bare approximation, which
takes into account only the lowest order corrections in $u$. Such
assumption is only true in the realm of weak-disorder
scattering.\cite{bruus} It can be seen that the renormalized spin
stiffness decreases linearly in this approximation. A naive
extrapolation of this dependence to the region of finite and
strong disorder strength shows that there is a certain value $u_p
= 0.7$, for which the renormalized spin stiffness vanishes (in the
long wavelength approximation $u_p = 1$), see
Fig.\;\ref{scottmurray3}(b). Green \cite{green} explains that a
vanishing renormalized spin stiffness at a threshold disorder
strength means that the 2DEG at $\nu = 1$ undergoes a quantum
phase transition from a ferromagnetic state to a paramagnetic one.
Thus, one can infer that the quantum Hall ferromagnet undergoes a
disorder-driven quantum phase transition to a paramagnetic state
at critical disorder strength $u_p = 0.7$. It is also interesting
to remark that Green established this general finding in the
domain of the weak disorder limit (though in the context of a
different model). The results obtained in this section cannot be
directly compared quantitatively with those of Green \cite{green},
Sinova \textit{et al.} \cite{macdgirvin} and Rapsch \textit{et
al.} \cite{chalkerlett} In addition to the fact that the model
used in the studies of Green is different, he does not complement
his proposition on the vanishing of the renormalized spin
stiffness with some quantitative results. Sinova \textit{et
al.}\cite{macdgirvin} use a disparate variable in the ratio of the
interaction strength to the Landau-level broadening disorder
energy scale. Finally, Rapsch \textit{et al.}\cite{chalkerlett}
perform their numerical calculations on a semiclassical spin
model.

In the next section we evaluate the self-energy using the
so-called self-consistent approximation and show that the
renormalized spin stiffness drastically changes its behavior,
which leads to completely different conclusions about the phase
transition.

\section{Self-consistent approximation}
The self-consistent approximation means that the self-energy is
evaluated with the total disorder averaged Green's function
(\ref{berabatamanutd4}) instead of the bare one.
Therefore, one has (see Eq.\;(\ref{fghw4})) 
\begin{equation}
\Sigma_u (p, \omega) = \frac{u}{4} \epsilon_B^2 e^{-p^2 /
2}\int^{\infty}_0 \ud q \ q e^{-q^2 / 2}[I_0 (q p) -1] \langle G_u
(q, \omega )\rangle_{\rm imp},
\end{equation}
Now, by referring to the computations carried out in the previous
section and substituting Eq.\;(\ref{berabatamanutd4}), one gets
\begin{equation}
\Sigma_u (p, \omega) = \frac{u}{4} \epsilon_B^2 e^{- p^2 / 2}
\int^{\infty}_0 \ud q \ q e^{- q^2 / 2} \frac{I_0 (q p)-1 }{\omega
- \omega_q - \Sigma_u (q, \omega) }.
\end{equation}
Using that
\begin{equation}
I_0 (q p)-1=\sum_{n=1}^{\infty}\frac{(q p)^{2n}}{(2^n n!)^2},
\end{equation}
one has
\begin{equation}
\Sigma_u (p, \omega) = \frac{u}{4} \epsilon_B^2 e^{- p^2 /
2}\sum_{n=1}^{\infty}\frac{p^{2n}}{(2^n n!)^2}\int^{\infty}_0 \ud
q \  \frac{q^{2n+1} e^{- q^2 / 2} }{\omega - \omega_{q} - \Sigma_u
(q, \omega) }.
\end{equation}
Thus, one can write
\begin{equation}
\Sigma_u (p, \omega) = e^{- p^2 /
2}\sum_{n=1}^{\infty}\sigma_n(\omega,u)p^{2n},
\end{equation}
with
\begin{equation}
\sigma_n(\omega,u) = \frac{u}{4(2^n n!)^2} \epsilon_B^2
\int^{\infty}_0 \ud q \  \frac{q^{2n+1} e^{- q^2 / 2} }{\omega -
\omega_{q} - \Sigma_u (q, \omega) }.
\end{equation}
Using such expansion allows one to promptly get a numerical
solution by iterations (see Fig.\;\ref{diaag11}).
\begin{figure}[h!]
\centering
\includegraphics[scale= 1]{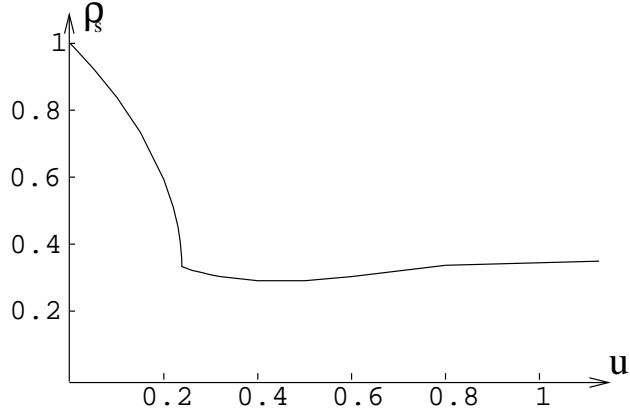}
\caption{Real part of the renormalized spin stiffness as a
function of the disorder strength $u$ in units of $\epsilon_B /
4$.} \label{diaag11}
\end{figure}
The convergence of the iterative solution is rather good up to
some value of the disorder concentration $u_c$. However, when
$u\rightarrow u_c$, we find that
$\partial_u\sigma_1(0,u)\rightarrow \infty$.
Therefore, it would be desirable to derive an analytical solution
in the neighborhood of $u_c$. For convenience, we omit the
arguments of $\sigma_n$ in our notation in the next part. In
general,
\begin{equation}\label{selfderiv}
\frac{\partial\Sigma_u (p, \omega)}{\partial u} =\frac{\Sigma_u (p,
\omega)}{u} +\frac{u}{4} \epsilon_B^2 e^{- p^2 / 2} \int^{\infty}_0
\ud q \ q e^{- q^2 / 2} \frac{ I_0 (q p) -1}{(\omega - \omega_{q} -
\Sigma_u (q, \omega))^2 }\frac{\partial\Sigma_u (q,
\omega)}{\partial u},
\end{equation}
or equivalently
\begin{equation}
\frac{\partial\sigma_n}{\partial u} =
\frac{\sigma_n}{u}+\frac{u}{4(2^n n!)^2} \epsilon_B^2
\int^{\infty}_0 \ud q \  \frac{q^{2n+1} e^{- q^2 / 2} }{(\omega -
\omega_{q} - \Sigma_u (q, \omega))^2 }\frac{\partial\Sigma_u (q,
\omega)}{\partial u}.
\end{equation}
\end{widetext}
 Introducing for simplicity
\begin{equation}\label{Fndef}
F_n \equiv \epsilon_B^2 \int^{\infty}_0 \ud q \  \frac{q^{2n+1} e^{-
q^2} }{[\omega - \omega_{q} - \Sigma_u (q, \omega)]^2}
\end{equation}
one finds
\begin{equation}\label{matreq}
\frac{\partial\sigma_n}{\partial u} =
\frac{\sigma_n}{u}+\frac{u}{4(2^n
n!)^2}\sum_{k=1}^{\infty}F_{n+k}\frac{\partial\sigma_k}{\partial
u}.
\end{equation}
Introducing a matrix notation
\begin{equation}\label{matr}
B_{m,n}\equiv \delta_{m,n}-\frac{uF_{m+n}}{2^{m+n+2} m!n!},
\end{equation}
Eq.\;\eqref{matreq} reads
\begin{equation}
\sum_{k=1}^{\infty}B_{n,k}2^k k!\frac{\partial\sigma_k}{\partial
u}=\frac{2^{n} n!\sigma_n}{u}.
\end{equation}
Its solution is found by computing the inverse matrix to
Eq.\;\eqref{matr} and has the form
\begin{equation}\label{derivu}
\frac{\partial\sigma_n}{\partial u}=\frac{2^{-n}}{u
n!}\sum_{k=1}^{\infty}B^{-1}_{n,k}2^{k} k!\sigma_k.
\end{equation}
Substituting this result into
\begin{equation}\nonumber
\frac{\ud (\det(B)^2)}{\ud
u}=2\det(B)\left(\frac{\partial\det(B)}{\partial
u}+\sum_{n=1}^{\infty}\frac{\partial\det(B)}{\partial\sigma_n}\frac{\partial\sigma_n}{\partial
u}\right),
\end{equation}
yields
\begin{equation}\label{detkappau}
\frac{\ud (\det(B)^2)}{\ud u}=-\kappa(u),
\end{equation}
where
\begin{eqnarray}\label{kappau}\nonumber
\kappa(u)&&\equiv -2\det(B)\frac{\partial\det(B)}{\partial
u}\\\nonumber
&&-2\det(B)\sum_{n,k=1}^{\infty}\frac{\partial\det(B)}{\partial\sigma_n}\frac{2^{k-n}k!}{u
n!}\sigma_k B^{-1}_{n,k},
\end{eqnarray} with
\begin{eqnarray}\nonumber
\frac{\partial\det(B)}{\partial u}&&=\det(B){\rm
Tr}\left(B^{-1}\frac{\partial B}{\partial u}\right)\\\nonumber
&&=-\frac{1}{u}\det(B){\rm Tr}(B^{-1}-I)
\end{eqnarray} and
\begin{equation}\nonumber
\frac{\partial\det(B)}{\partial\sigma_n}=
-\sum_{m,k=1}^{\infty}\det(B)B^{-1}_{k,m}\frac{u}{2^{m+k+2}
m!k!}\frac{\partial F_{m+k}}{\partial\sigma_n},
\end{equation}
where
\begin{equation}\nonumber
\frac{\partial F_{m}}{\partial\sigma_n} = 2\epsilon_B^2
\int^{\infty}_0 \ud q \  \frac{q^{2n+2m+1} e^{- 3q^2/2} }{[\omega
- \omega_{q} - \Sigma (q, \omega)]^3}.
\end{equation}
 Suppose that $\det(B) \rightarrow 0$ when $u \rightarrow
u_c$. In this case $ B^{-1}\det (B)$ remains finite, as well as
$\kappa(u)$. This suggests that
\begin{equation}
\frac{\partial\sigma_n}{\partial u}\rightarrow \infty.
\end{equation}
when $u \rightarrow u_c$, since the other terms are finite.
Moreover, if $\kappa(u)$ is a smooth function around $u_c$, such
that $\kappa(u_c)\approx \kappa(u_0)$ for some $u_0$ from the
neighborhood of $u_c$, then according to Eq.\;\eqref{detkappau}
there holds
\begin{equation}\label{det}
\det[B(u)]=\sqrt{\kappa(u_c)(u_c-u)}+{\cal O}(u_c-u).
\end{equation}
It follows then from Eq.\;\eqref{det} that $u_c\approx
u_0+\det[B(u_0)]^2/\kappa(u_0)$ as long as $u_0\rightarrow u_c$.
However, the analysis of the infinite dimensional matrix $B$ and its
determinant is quite complicated, which forces us to use an
approximate solution, where we keep only the first $40$ terms in the
expansion, thus reducing the dimension of the matrices to
$40\times40$. In the absence of Zeeman splitting ($g=0$), for
$\omega=0$, and $u_0=0.238$ one finds, setting $\epsilon_B=1$, that
$\det[B(u_0)]=0.0551776$ and $\kappa(u_0)=9.7945$, which yields
$u_c=0.238311$ in excellent agreement with the numerical solution.
The approximation also allows to check the validity of
Eq.\;\eqref{derivu}, which yields $\sigma'_1=-9.384$ at the point
$u_0=0.238$ (here the prime stands for the partial derivative with
respect to $u$). On the other hand, the numerical solution for the
two points $u_0=0.238$ and $u_1=0.23801$ yields $\Delta
\sigma_1/\Delta u=-9.463$, which agrees reasonably well with the
previous result. The main difference stems from the fact that
$u_0=0.238$ is rather close to the critical point $u_c$, where the
derivative diverges, so the value $\Delta u=10^{-5}$ is still rather
large and, of course, computational errors and approximation with
finite number of terms make the result not very precise.
Furthermore, it follows from Eq.\;\eqref{derivu} that $\sigma'_n
\det(B)$ remains finite with $u \rightarrow u_c$. Thus,
\begin{eqnarray}\nonumber
&&\sigma_n(u)-\sigma_n(u_0)=\int^{u}_{u_0} \ud
v\frac{\partial\sigma_n(v)}{\partial v}\\
&&\approx\frac{\det[B(u_0)]\sigma'_n(u_0)}{\sqrt{\kappa(u_0)}}\int^{u}_{u_0}
\ud v(u_c-v)^{-1/2},
\end{eqnarray}
which leads to
\begin{equation}
\sigma_n(u)=\sigma_n(u_c)+[\sigma_n(u_0)-\sigma_n(u_c)]\frac{\det[B(u)]}{\det[B(u_0)]}
\end{equation}
after performing the integration, where
$\sigma_n(u_c)-\sigma_n(u_0)=2(u_c-u_0)\sigma'_n(u_0)$. From this
analytic solution one may observe that $\sigma_n$ and,
consequently, $\Sigma (\textbf{p}, \omega)$ acquires an imaginary
part when $u>u_c$. In particular, considering $n=1$, for the case
at hand $\sigma_1(u_0)=-0.161742$ and $\sigma_1(u_c)=-0.167576$.
Defining
\begin{equation}
\alpha\equiv-\lim_{u\rightarrow u_c}\frac{2\sigma'_1(u)
\det[B(u)]}{\sqrt{\kappa(u)}},
\end{equation}
the value of $\alpha$ can be evaluated without any fitting
parameters directly from Eqs.\; \eqref{derivu} and \eqref{kappau},
which yields $\alpha=0.331$. It follows directly from the above
that the renormalized spin stiffness now obeys
\begin{equation}
\rho_s(u) =\rho_s(u_c)+ \alpha\epsilon_B\sqrt{u_c-u},
\end{equation}
where $\rho_s(u_c)=\epsilon_B(\sigma_1(u_c)+1/4)$. Both numerical
and analytic results for $\rho_s(u)$ are plotted in
Fig.\;\ref{scottmurray5}, which shows that the analytic solution
remains in excellent agreement with the numerical one even for
those values of $u$, which are far from the critical point $u_c$.
The whole behavior of the renormalized spin stiffness is very
similar to the one obtained by Chalker \textit{et
al.},\cite{chalkerlett} describing a spin glass phase transition.
Moreover, such dependence of the renormalized spin stiffness as a
square root function of a control parameter was already observed
previously by Shender,\cite{Shender} as well as by Avgin
\textit{et al.}\cite{Avgin} They considered the two- and tree-
dimensional $\pm J$ Heisenberg spin glass model in a ferromagnetic
ground state due to a strong external magnetic field. They found
that for a certain value of the control parameter, $\rho_s(u)$
acquires an imaginary part. The real part of $\rho_s(u)$ is
proportional to the spin wave stiffness, whereas the imaginary
part is proportional to the damping of the spin wave excitations,
thus signalling localization. It was argued that when the
frequency of the spin-wave excitation $\omega$ multiplied by its
lifetime $\tau$ is $\omega\tau={\rm Re}[\rho_s(u)]/{\rm
Im}[\rho_s(u)]<1$, then the spin waves are completely localized.
As we can see from the Fig.\;\ref{scottmurray5}, the condition of
localization is already satisfied for the values of the disorder
strength starting from $u=0.3$. The calculations presented in the
Appendix D contain a strong indication that the Pauli
susceptibility diverge at the point $u=u_c$, suggesting a phase
transition from a ferromagnetic ground state to a spin glass
state,\cite{chalkerlett} since the spin waves become localized.

 Our discussion was mainly concerned with the static
case $\omega=0$. However, our approach allows to find $\Sigma (p,
\omega)$ for any given $\omega$. The dispersion spectrum in the
self-consistent approximation then satisfies $ \omega - \omega_{p} -
\text{Re} \Sigma (p, \omega) = 0 $.
\begin{figure}[t!]
\centering
\includegraphics[scale=0.87]{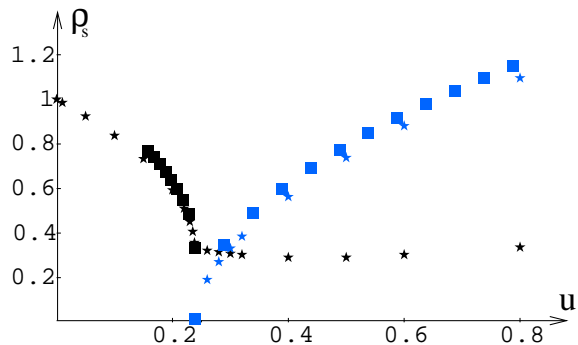}
\caption{(Color online) Real (black) and imaginary (blue) parts of
the renormalized spin stiffness in units of $\epsilon_B / 4$ as
functions of the disorder strength $u$. (Square - analytic
solution, star - numerical)}
\label{scottmurray5}
\end{figure}


\section{Conclusions \& Outlook}

This paper accounts for the presence of both disorder and
interactions in a 2DEG at Landau level filling factor $\nu = 1$,
whose ground state constitutes the well-known quantum Hall
ferromagnet. The bosonization technique developed by Doretto
\textit{et al.} \cite{doretto} was employed in order to facilitate
the treatment of both disorder and interactions in this strongly
correlated system. The bosonization procedure consists in treating
the spin wave (magnon) excitation as a boson such that the
fermionic Hamiltonian of the system can be approximately recast
into a Hamiltonian expressed in terms of bosonic operators. As a
consequence, the interaction between electrons up to RPA level was
incorporated within the bare propagator that represents the free
boson. The intent was then to identify a disorder-driven quantum
phase transition to a non-ferromagnetic state by analyzing the
behavior of the renormalized spin stiffness as a function of the
disorder strength, which itself corresponds to the ratio squared
of the disorder energy scale to the Coulomb energy one. To achieve
this aim, firstly, we derived the bosonic expression for the
Hamiltonian of the system. In the second stage, the focus was on
seeking out the disorder self-averaged Green's function, which is
the full bosonic Green's function averaged over the impurity
positions. Then, by using the Dyson's equation, we obtained a
diagrammatic representation of the self-energy. The latter was
subsequently computed within the framework of the low-density
weak-scattering approximation. Low density means that the number
of disorder atoms present in the system is taken to be much lower
than the number of electrons, while the weak-scattering
approximation signifies that the scattering potential induced by a
given impurity atom is weak, such that only the first and
second-order Born scatterings are accounted for. As a result, the
self-energy corresponded to a single diagram. Furthermore, the
self-energy was evaluated in three different approximations: 1)
the bare (long wavelength) approximation, which consists in using
the bare bosonic propagator and keeping the lowest order terms in
momenta, 2) the bare (full $k$) approximation, which uses as well
the bare bosonic propagator but with all the momenta terms kept in
the calculation and, finally, 3) the self-consistent
approximation, which uses the full disorder averaged Green's
function instead of the bare one in the self-energy diagram. Then,
the renormalized spin stiffness was determined by extracting the
coefficient of the quadratic term in the dispersion relation
together with the contribution from the self-energy. In the case
of the bare (long wavelength) approximation, the spin stiffness
was found to vanish linearly at the disorder strength $u_p = 1$.
For the bare (full $k$) scheme, the spin stiffness also vanished
linearly, but at the disorder strength $u_p = 0.7$. These results
suggest the occurrence of a disorder-driven quantum phase
transition from the ferromagnetic phase to a paramagnetic one at
the critical value $u_p = 0.7$. Lastly, the self-consistent
calculation revealed a completely different behavior: the real
part of the renormalized spin stiffness also initially decreases
with increasing the disorder strength $u$, but then it saturates
without reaching zero beyond a critical value $u_c$, at which it
(and the self-energy) acquires an imaginary component. According
to the Shender criterium,\cite{Shender} the spin waves become
completely localized when the imaginary part of the renormalized
spin stiffness becomes larger than the real part, which occurs in
our system for $u>0.3$ (see Fig.\;\ref{scottmurray5}).

The physical mechanism behind a phase transition from the
ferromagnetic ground state can be understood by considering
electrons completely filling the lowest Landau level ($\nu = 1$) in
the presence of some inhomogeneous electrostatic background
(disorder). Then, for sufficiently strong impurity potential, by
adjusting the electron density to the electrostatic background, the
system would gain more energy than is needed to rearrange the spin
configuration. In this case the ferromagnetic state does not
minimize the total energy of the system and a phase transition
should take place. This quantum phase transition could be detected
by calculating the behavior of the magnetic susceptibility as a
function of the disorder strength. A sharp peak is anticipated at
the transition point. In particular, if the energy cost for exciting
a spin wave is less than the gain in the electrostatic energy, then
the renormalized spin stiffness becomes negative and the system
undergoes a phase transition to a paramagnetic state with zero local
magnetization. On the other hand, as it was argued by Rapsch
\textit{et al.}, \cite{chalkerlett} in the case of a smoothly
varying impurity potential, keeping nonzero local magnetization is
still energetically favorable and the electrostatic energy is
lowered by screening the impurity potential due to the formation of
spin textures. At strong disorder such phase would correspond to a
spin glass and the spin textures might be considered as the
localized spin waves. Thus, the character of the phase transition
might depend on the nature of the disorder. The calculations
performed within our model indicate that the Pauli susceptibility
diverges at the same critical point of the disorder strength $u_c$,
where an imaginary part of the renormalized spin stiffness appears,
thus suggesting a phase transition to a spin glass phase.

Our approach can be extended for the case of bilayer systems in
the presence of disorder. In fact, Fertig and Murthy
\cite{fertigmurthy} have already considered such systems. Thus, it
would be interesting to apply our formalism to the case of a
bilayer system with the total filling factor $\nu_T = 1$ and
compare the results.

\section{Acknowledgments}
We acknowledge insightful discussions with T.\ Giamarchi and L.\
Cugliandolo. We also would like to thank R.\ L.\ Doretto for
proposing us this interesting problem. This work was partially
supported by the Netherlands Organization for Scientific Research
(NWO).
\appendix
\begin{widetext}
\section{Third order diagram}

Let us now evaluate the $3^{\text{rd}}$ order diagram contribution
to the self-energy. Its diagrammatic representation is shown in
Fig.\ 6.
Algebraically, we have
\begin{equation}
\begin{split}
\Sigma^{(3)} (\textbf{p}, \omega) &  = N_{\rm imp} \sum_{\textbf{q},
\textbf{q}'} U^{e} (\textbf{q} - \textbf{p}, \textbf{p}) G
(\textbf{q}, \omega) U^{e} (\textbf{q}' - \textbf{q}, \textbf{q}) G
(\textbf{q}', \omega)
U^{e}  (\textbf{p} - \textbf{q}', \textbf{q}' ) \\
& = N_{\rm imp} \sum_{\textbf{q}, \textbf{q}'} U (\textbf{q} -
\textbf{p}) f(\textbf{q} - \textbf{p},\textbf{p}) G (\textbf{q},
\omega) U(\textbf{q}'
- \textbf{q}) f(\textbf{q}' - \textbf{q}, \textbf{q}) \\
& \qquad \qquad \times G (\textbf{q}', \omega) U (\textbf{p} -
\textbf{q}') f(\textbf{p} - \textbf{q}', \textbf{q}' ).
\end{split}
\end{equation}
\begin{figure}[h!]
\centering
\includegraphics[scale= 0.4]{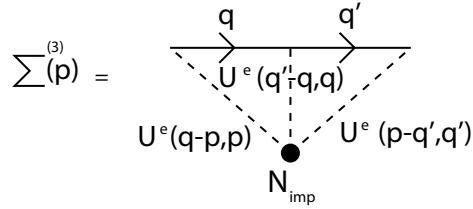}
\label{diaag12} \caption{Diagrammatic representation of the third
order diagrammatic contribution to the self-energy.}
\end{figure}
Here again, the impurity potential is short-range $U (\textbf{q})
= \text{constant} = U / A$. By replacing all the functions defined
previously, we find
\begin{equation}
\begin{split}
\Sigma^{(3)} (\textbf{p}, \omega) &  = N_{\rm imp}
\sum_{\textbf{q}, \textbf{q}'} \left( \frac{U}{A} \right)^3 \left[
2 i e^{- |\ell (\textbf{q} - \textbf{p})|^2 / 4} \sin \left(
\frac{(\textbf{q} - \textbf{p})
\wedge \textbf{p}}{2} \right) \right] G (\textbf{q}, \omega) \\
& \qquad \times \left[ 2 i e^{- |\ell (\textbf{q}' -
\textbf{q})|^2 / 4} \sin \left( \frac{(\textbf{q}' - \textbf{q})
\wedge \textbf{q}}{2} \right) \right] G (\textbf{q}', \omega)
\left[ 2 i e^{- |\ell (\textbf{p} - \textbf{q}')|^2 / 4}
\sin \left( \frac{(\textbf{p} - \textbf{q}') \wedge \textbf{q}'}{2} \right) \right] \\
& = N_{\rm imp} \sum_{\textbf{q}, \textbf{q}'} \left( \frac{U}{A}
\right)^3  (2 i)^3 \underset{\textbf{Term a}}{\underbrace{ e^{-
|\ell (\textbf{q} - \textbf{p})|^2 / 4} e^{- |\ell (\textbf{p} -
\textbf{q}')|^2 / 4} e^{- |\ell (\textbf{q}' - \textbf{q})|^2 / 4}
}} \underset{\textbf{Term b}}{\underbrace{\sin \left(
\frac{\textbf{q} \wedge \textbf{p}}{2} \right)  \sin \left(
\frac{\textbf{p} \wedge \textbf{q}'}{2}
\right) }} \\
& \qquad \times \underset{\textbf{Term c}}{\underbrace{ \sin
\left( \frac{\textbf{q}' \wedge \textbf{q}}{2} \right) }} G
(\textbf{q}, \omega) G (\textbf{q}', \omega).
\end{split}
\end{equation}
It is clear that Terms a and b are symmetric under the interchange
$\textbf{q} \leftrightarrow \textbf{q}'$ while Term c is
antisymmetric. Thus, one has
\begin{equation}
\Sigma^{(3)} (\textbf{p}, \omega) = 0
\end{equation}
This result holds true in both the full bare and self-consistent
approximations.

As a matter of fact, due to the antisymmetric property of the
wedge product within the sine term, it turns out that all odd
order terms vanish.

\section{Detailed derivation of the self-energy}

We prove here the expressions \eqref{saaidismybro2} and
\eqref{saaidismybro2bisbis}. We begin with the expression of the
self-energy given by Eq.\;(\ref{fghw2})
\begin{equation}
\Sigma (p, \omega) = \ 4 n_{imp} U^2 \int^{\infty}_0 \frac{\ud
q}{2 \pi} \ q e^{-(\ell q)^2 /2} e^{-(\ell p)^2 /2} G (q, \omega )
\int^{2 \pi}_0 \frac{\ud \phi }{2 \pi} \ e^{\ell^2 q p \cos \phi}
\frac{1}{2} [1 - \cos (\ell^2 q p \sin \phi) ]. \label{a1}
\end{equation}
One first deals with the polar integral,
\begin{equation}\label{ghigbigb3}
\int^{2 \pi}_0 \frac{\ud \phi }{2 \pi} \ e^{\ell^2 q p \cos \phi} \frac{1}{2} [1 -
\cos (\ell^2 q p \sin \phi) ] = \frac{1}{2} \int^{2 \pi}_0 \frac{\ud \phi }{2 \pi}
\ e^{\ell^2 q p \cos \phi} - \frac{1}{2}  \int^{2 \pi}_0 \frac{\ud \phi }{2 \pi}
\ e^{\ell^2 q p \cos \phi} \cos (\ell^2 q p \sin \phi).
\end{equation}
The two terms are evaluated separately. For the first term, one must note that
\cite{gradshteyn}
\begin{equation}\nonumber
e^{\ell^2 q p \cos \phi} = I_0 (\ell^2 q p) + 2 \sum^{\infty}_{n = 1} I_n (\ell^2 q p) \cos (n \phi),
\end{equation}
such that
\begin{equation}\label{ghigbigb1}
\frac{1}{2} \int^{2 \pi}_0 \frac{\ud \phi }{2 \pi} \ e^{\ell^2 q p \cos \phi} = \frac{1}{2} I_0 (\ell^2 q p) \int^{2 \pi}_0 \frac{\ud \phi }{2 \pi} + 2 \sum^{\infty}_{n = 1} I_n (\ell^2 q p) \frac{1}{2} \int^{2 \pi}_0 \frac{\ud \phi }{2 \pi} \ \cos (n \phi) = \frac{1}{2} I_0 (\ell^2 q p).
\end{equation}
The second term 

\begin{equation}\label{ghigbigb2}
\frac{1}{2}  \int^{2 \pi}_0 \frac{\ud \phi }{2 \pi} \ e^{\ell^2 q
p \cos \phi} \cos (\ell^2 q p \sin \phi)=\frac{1}{2}  \int^{2
\pi}_0 \frac{\ud \phi }{2 \pi} \ \exp[{\ell^2 q p \cdot
\exp(i\phi)}]= \frac{1}{2}
\end{equation}
Substituting Eqs.\;(\ref{ghigbigb1}) and (\ref{ghigbigb2}) back
into Eq.\;(\ref{ghigbigb3}) then yields the simpler expression,
\begin{equation}\nonumber
\int^{2 \pi}_0 \frac{\ud \phi }{2 \pi} \ e^{\ell^2 q p \cos \phi}
\frac{1}{2} [1 - \cos (\ell^2 q p \sin \phi) ] = \frac{1}{2}[ I_0
(\ell^2 q p)-1] .
\end{equation}
Now, substituting the above back into Eq.\;(\ref{a1}) we find
\begin{equation}\label{fghw4b}
\Sigma (p, \omega) = 4 n_{\rm imp} U^2 \int^{\infty}_0 \frac{\ud
q}{2 \pi} \ q e^{-(\ell q)^2 /2} e^{-(\ell p)^2 /2} G (q, \omega )
\frac{1}{2}[ I_0 (\ell^2 q p)-1].
\end{equation}

\section{Bare (Long Wavelength) Approximation}
To evaluate the self-energy within the long-wavelength
approximation, we must return to Eq.\;(\ref{fghw1}). Firstly, one
remarks that the sine squared term in Eq.\;(\ref{fghw1}) greatly
simplifies,
\begin{equation}\label{longsine}
\sin^2 \left( \frac{\textbf{q} \wedge \textbf{p}}{2} \right) \approx
\left( \frac{\textbf{q} \wedge \textbf{p}}{2} \right)^2  =
\frac{1}{4}  [ \ell^2 \hat{z} \cdot (\textbf{q} \times
\textbf{p})]^2 = \frac{1}{4} ( \ell^2 |\textbf{q} \times
\textbf{p}|)^2 = \frac{1}{4} \ell^4 |\textbf{q}|^2 |\textbf{p}|^2
\sin^2 \phi.
\end{equation}
Then, substituting Eq.\;(\ref{longsine}) into Eq.\;(\ref{fghw1})
yields:
\begin{equation}\label{stefanos1}
\Sigma (\textbf{p}, \omega) = n_{\rm imp} U^2 \ell^4
\int^{\infty}_0 \frac{\ud q}{2 \pi} \ q^3 e^{- (\ell q)^2 / 2} p^2
e^{- (\ell p)^2 / 2} G^0 (\textbf{q}, \omega) \int^{2 \pi}_0
\frac{\ud \phi}{2 \pi} \ e^{\ell^2 p q \cos \phi} \sin^2 \phi.
\end{equation}
\end{widetext}
The polar integral then turns out to match \cite{gradshteyn}
\begin{equation}\label{IntBessel}
\int^{2 \pi}_0 \frac{\ud \phi}{2 \pi} \ e^{\ell^2 p q \cos \phi}
\sin^2 \phi = \frac{1}{2} [I_0 (\ell^2 p q) - I_2 (\ell^2 p q)].
\end{equation}
The series expansion for Eq.\;(\ref{IntBessel}) gives ($x \equiv
\ell^2 p q$)
\begin{equation}
\begin{split}
I_0 (x) - I_2 (x) & = \left[ 1 + \frac{x^2 }{4} + \dots \right] -
\left[ \frac{x^2 }{8} + \ldots \right] \\ & = 1 + \frac{x^2}{8} +
\ldots
\end{split}
\end{equation}
Eq.\;(\ref{stefanos1}) already holds a $q^3 p^2$ term and
therefore a $q^5 p^4$ term is not needed in the long wavelength
approximation. Thus, one assumes that
\begin{equation}
I_0 (\ell^2 p q) - I_2 (\ell^2 p q) \ \approx \ 1.
\end{equation}
Moreover, the momenta are rescaled as $\textbf{q}, \textbf{p}
\rightarrow \textbf{q} / \ell , \textbf{p} / \ell$. As a result,
Eq.\;(\ref{stefanos1}) simplifies to
\begin{equation}\nonumber
\Sigma (p, \omega) \  = \frac{n_{\rm imp} U^2}{4 \pi \ell^2} p^2
e^{- p^2 / 2} \int^{\infty}_0 \ud q \ q^3 e^{-q^2 / 2}
\frac{1}{\omega - \omega_{q} + i \eta},
\end{equation}
where we replaced $G^0 (\textbf{q} , \omega)$ by its definition
(see Eq.\;\eqref{berbatovatmanutd1}). One can then make use of the
identity \cite{doniach}
\begin{equation}
\frac{1}{x + i \eta} = \mathcal{P} \frac{1}{x} - i \pi \delta (x),
\end{equation}
where $\mathcal{P}$ symbolizes the Cauchy principal value of the
integral. Consequently, one has
\begin{align}
\text{Re} \Sigma (p, \omega)  & = \left( \frac{\epsilon_B}{4}
\right)^2 u \ p^2 e^{- p^2 / 2} \mathcal{P} \int^{\infty}_0 \ud q
\frac{q^3 e^{-q^2 / 2} }{\omega
- \omega_{q}}, \\
\text{Im} \Sigma (p, \omega)  & = - \left( \frac{\epsilon_B}{4}
\right)^2 u \ p^2
e^{- p^2 / 2} \int^{\infty}_0 \ud q \ q^3 e^{-q^2 / 2} \nonumber \\
& \qquad \qquad \qquad \qquad \qquad \quad \times \pi \delta
(\omega - \omega_{q}) .\label{gmailrocks1}
\end{align}

Let us first examine the real part of the self-energy, which
actually denotes the physical self-energy.

It has been shown\cite{doretto} that in the long wavelength
approximation the bosonic dispersion relation for $\textbf{q}$ can
be written as
\begin{equation}
\omega_{\textbf{q}} = g + \frac{\epsilon_B}{4} q^2.
\end{equation}
The physical self-energy then becomes
\begin{equation}
\text{Re} \Sigma (p, \omega) = \left( \frac{\epsilon_B}{4}
\right)^2 u \ p^2 e^{- p^2 / 2} \mathcal{P} \int^{\infty}_0 \ud q
\frac{q^3 e^{-q^2 / 2} }{\omega - g - \frac{\epsilon_B}{4} q^2}.
\end{equation}
Let us then work temporarily with the new quantities
\begin{equation}
\bar{\omega} = \frac{4 \omega}{ \epsilon_B} \ \ \text{and} \ \
\bar{g} = \frac{4 g}{\epsilon_B},
\end{equation}
such that the self-energy is re-written as
\begin{equation}\nonumber
\text{Re} \Sigma (p, \bar{\omega}) = \frac{\epsilon_B}{4} u \ p^2
e^{- p^2 / 2} \mathcal{P} \int^{\infty}_0 \ud q \frac{q^3 e^{-q^2
/ 2}}{\bar{\omega} - \bar{g} - q^2}.
\end{equation}
Now, one performs a change of variable in the $q$ momentum: $q
\rightarrow \tilde{q} = q^2$. One must note that $q dq = d (q^2) /
2$ and that the integration limits are not altered. Consequently,
one gets
\begin{equation}
\text{Re} \Sigma (p, \bar{\omega}) = \frac{\epsilon_B}{4} u \
 p^2 e^{- p^2 / 2} \mathcal{P} \int^{\infty}_0
\frac{\ud \tilde{q}}{2} \frac{\tilde{q} e^{- \tilde{q} /
2}}{\bar{\omega} - \bar{g} - \tilde{q}}.
\end{equation}
A further change of the integration variable is performed
$\tilde{q} \rightarrow k = \bar{\omega} - \bar{g} - \tilde{q}$,
leading to
\begin{widetext}
\begin{equation}
\begin{aligned}
\text{Re} \Sigma (p, \bar{\omega}) &  = - \frac{\epsilon_B}{4} u \
p^2 e^{- p^2 / 2} \mathcal{P} \int^{- \infty}_{\bar{\omega} -
\bar{g}} \frac{\ud k}{2} \ (\bar{\omega} - \bar{g} - k) e^{-
(\bar{\omega}
- \bar{g} - k) / 2} \frac{1}{k} \\
& = \frac{\epsilon_B}{4} u \ p^2 e^{- p^2 / 2} \left[ \left(
\frac{\bar{\omega} - \bar{g}}{2} \right) \underset{\textbf{Term
a}}{\underbrace{\left( \mathcal{P} \int^{\bar{\omega} -
\bar{g}}_{- \infty} \ud \left( \frac{k}{2} \right) \ \frac{e^{( k
/ 2 )} }{\left( \frac{k}{2} \right) } \right) }} e^{- (
\bar{\omega}
- \bar{g} ) / 2} \right. \\
& \qquad \qquad \qquad \qquad \qquad \qquad \qquad - \left.
\frac{1}{2} \underset{\textbf{Term b} } {\underbrace{\left(
\mathcal{P} \int^{\bar{\omega} - \bar{g} }_{- \infty} \ud k \ k
\frac{e^{k / 2}}{k} \right) }} e^{- ( \bar{\omega} - \bar{g} ) /
2} \right] .
\end{aligned}
\end{equation}
\end{widetext}
Term a corresponds to the definition of the exponential integral
function;\cite{gradshteyn}
\begin{equation}
\text{Ei} \left( \frac{\bar{\omega} - \bar{g}}{2} \right) =
\mathcal{P} \int^{\bar{\omega} - \bar{g}}_{- \infty} \ud \left(
\frac{k}{2} \right) \ \frac{e^{( k / 2 )}}{\left( \frac{k}{2}
\right) },
\end{equation}
whereas Term b can be straightforwardly integrated,
\begin{equation}
\begin{split}
\mathcal{P} \int^{\bar{\omega} - \bar{g}}_{- \infty} \ud k \ k
\frac{e^{k / 2}}{k} & = \int^0_{- \infty} \ud \tilde{k} \
e^{(\tilde{k} + \bar{\omega} - \bar{g}) / 2}
\\
& = \left( \int^0_{- \infty} \ud \tilde{k} \ e^{\tilde{k}} \right)
e^{(\bar{\omega} - \bar{g}) / 2} = 2 e^{(\bar{\omega} - \bar{g}) /
2},
\end{split}
\end{equation}
where the shift of variable $k \rightarrow \tilde{k} = k -
(\bar{\omega} - \bar{g})$ was used in the first step.

Thus, the physical self-energy becomes
\begin{equation}
\begin{split}
\text{Re} \Sigma (p, \bar{\omega}) & = \frac{\epsilon_B}{4} u \
p^2 e^{- p^2 / 2} \left[ -1 \right.
\\
& + \left. \left( \frac{\bar{\omega} - \bar{g} }{2} \right)
\text{Ei} \left( \frac{\bar{\omega} - \bar{g}} {2} \right) e^{- (
\bar{\omega} - \bar{g} ) / 2} \right].
\end{split}
\end{equation}


The renormalized energy of the bosons is obtained by looking at
the poles of the full disorder self-averaged Green's function,
\begin{equation}\label{saadismybro4}
\omega - \omega_{\textbf{p}} - \text{Re} \Sigma (\textbf{p},
\omega) = 0.
\end{equation}
Consequently, in the long wavelength approximation, the
renormalized dispersion relation takes the form
\begin{equation}
\begin{split}
\bar{\omega} - \bar{g} & = p^2 + u
\ p^2 e^{- p^2 / 2} \left[ -1 \right. \\
& + \left. \left( \frac{\bar{\omega} - \bar{g}}{2} \right) \text{Ei}
\left( \frac{\bar{\omega} - \bar{g}}{2} \right) e^{- ( \bar{\omega}
- \bar{g} ) / 2} \right].
\end{split}
\end{equation}

It is straightforward to notice that the renormalized spin
stiffness, which corresponds to the coefficient of the $p^2$ term,
is given by
\begin{equation}
\rho_s^R = \frac{\epsilon_B}{4} (1 - u).
\end{equation}


We now turn to the imaginary part of the self-energy given by
Eq.\eqref{gmailrocks1}. In the long wavelength approximation, the
Dirac delta function becomes
\begin{equation}\label{berbaatmanutd2}
\delta ( \omega - \omega_{q} ) \approx \delta \left( \omega -
\left( g + \frac{\epsilon_B}{4} q^2 \right) \right) =
\frac{4}{\epsilon_B} \delta ( \bar{\omega} - \bar{g} - q^2 ).
\end{equation}
By performing a change of variable in the $q$ momentum, $q
\rightarrow \tilde{q} = q^2$ and replacing
Eq.\;\eqref{berbaatmanutd2} into Eq.\;\eqref{gmailrocks1} one gets
\begin{equation}
\text{Im} \Sigma (p, \bar{\omega}) = - \frac{\pi}{2} u \ p^2 e^{-
p^2 / 2} (\omega - g) e^{- 2 ( \omega - g ) / \epsilon_B}.
\end{equation}

Finally, the scattering time, which amounts to the lifetime of the
bosonic excitation, is given by
\begin{equation}
\frac{1}{\tau_{\textbf{p}}} = \pi u \ p^2 e^{- p^2 / 2} (\omega -
g) e^{- 2 ( \omega - g ) / \epsilon_B}.
\end{equation}
It is clear that $\tau_{\textbf{p}} \rightarrow \infty$ when
$\omega \rightarrow g$, i.e. low energy quasiparticles are
long-lived, with finite lifetime induced by disorder.

\section{Pauli susceptibility}
The Pauli susceptibility in case of linear response is given by
the Kubo formula:
\begin{equation}
\chi_{zz}(\textbf{x},\textbf{x}'; t-t')=i \langle T_t
S_z(\textbf{x}, t)S_z(\textbf{x}',t')\rangle.
\end{equation}
Using the Fourier transformation
\begin{equation}
S_z(\textbf{x},
t)=\sum_{\textbf{q}}S_z(\textbf{q},t)e^{i\textbf{q}\cdot\textbf{x}},
\end{equation}
the susceptibility can be written as
\begin{equation}
\chi_{zz}(\textbf{q},\textbf{q}'; t-t')=i\langle T_t
S_z(\textbf{q}, t)S_z(\textbf{q}',t')\rangle.
\end{equation}
On the other hand, the operators $S_z(\textbf{q}, t)$ can be written
in the bosonized form\cite{doretto}
\begin{equation}\nonumber
S_z(\textbf{q},
t)=\frac{N_\phi}{2}\delta_{\textbf{q},0}-e^{-\textbf{q}^2/4}\sum_{\textbf{p}}\cos\left(
\frac{\textbf{q} \wedge \textbf{p}}{2}
\right)b^{\dagger}_{\textbf{q}+\textbf{p}}(t)b_{\textbf{p}}(t),
\end{equation}
where $b_{\textbf{p}}(t)=e^{iHt}b_{\textbf{p}}e^{-iHt}$. Thus,
after substitution
\begin{eqnarray}\nonumber
\chi_{zz}(\textbf{q},\textbf{q}';
t-t')&=&ie^{-\textbf{q}^2/2}\sum_{\textbf{p},\textbf{p}'}\cos\left(
\frac{\textbf{q} \wedge \textbf{p}}{2} \right)\cos\left(
\frac{\textbf{q} \wedge \textbf{p}'}{2} \right)\\\nonumber
&\times&\langle T_t
b^{\dagger}_{\textbf{q}+\textbf{p}}(t)b_{\textbf{p}}(t)b^{\dagger}_{\textbf{q}'+\textbf{p}'}(t')b_{\textbf{p}'}(t')\rangle.
\end{eqnarray}
Evaluation of the expectation value yields
\begin{eqnarray}\nonumber
\chi_{zz}(\textbf{q},\textbf{q}';
t-t')=-ie^{-\textbf{q}^2/2}\sum_{\textbf{p},\textbf{p}'}\cos\left(
\frac{\textbf{q} \wedge \textbf{p}}{2} \right)\cos\left(
\frac{\textbf{q} \wedge \textbf{p}'}{2} \right)\\\nonumber \times
G(\textbf{p}'+\textbf{q}',\textbf{p};t-t')G(\textbf{p}+\textbf{q},\textbf{p}';t'-t),
\end{eqnarray}
using the notation defined earlier in Eq.\;(\ref{Greendoniach}).
Expanding the Green's function $G(\textbf{p},\textbf{q}; t-t')$
into the Born series and performing the disorder averaging one
recovers the translational invariance
$\langle\chi_{zz}(\textbf{q},\textbf{q}'; t-t')\rangle_{\rm
imp}=\delta_{\textbf{q}+\textbf{q}',0}\chi_{zz}(\textbf{q},
t-t')$. Moreover, performing the Fourier transformation in the
time variable $t$ and introducing
\begin{eqnarray}\nonumber
P(\textbf{p},\textbf{q};\omega,\epsilon)&\equiv&\sum_{\textbf{p}'}\cos\left(
\frac{\textbf{q} \wedge \textbf{p}'}{2} \right)\\ \nonumber
&\times&\langle G(\textbf{p}'-\textbf{q},\textbf{p};\omega+
\epsilon)G(\textbf{p}+\textbf{q},\textbf{p}';\omega)\rangle_{\rm
imp}
\end{eqnarray}
the susceptibility is

\begin{widetext}
\begin{equation}
\chi_{zz}(\textbf{q},
\epsilon)=\frac{-iAe^{-\textbf{q}^2/2}}{(2\pi)^3}\int \ud
\textbf{p}\int_{-\infty}^{\infty} \ud \omega \cos\left(
\frac{\textbf{q} \wedge \textbf{p}}{2}
\right)P(\textbf{p},\textbf{q};\omega, \epsilon).
\end{equation}
In the self consistent approximation the function
$P(\textbf{p},\textbf{p}';\omega,\epsilon)$ obeys\cite{doniach}
\begin{equation}\nonumber
P(\textbf{p},\textbf{p}';\omega,\epsilon)=G(\textbf{p},
\omega+\epsilon)G(\textbf{p}+\textbf{p}', \omega)\left[\cos\left(
\frac{\textbf{p} \wedge \textbf{p}'}{2} \right)+\frac{AN_{\rm
imp}}{(2\pi)^2}\int \ud \textbf{q}
U^e(\textbf{p}-\textbf{q},\textbf{q})U^e(\textbf{q}-\textbf{p},\textbf{p}+\textbf{p}')
P(\textbf{q},\textbf{p}';\omega,\epsilon)\right].
\end{equation}
We are interested mostly in the static susceptibility $\chi\equiv
\lim_{\epsilon\rightarrow 0} \chi_{zz}(0, \epsilon)$. Thus, in
particular
\begin{equation}\nonumber
P(\textbf{p},0;\omega,0)=G^2(\textbf{p},
\omega)\left[1+\frac{AN_{\rm imp}}{(2\pi)^2}\int \ud \textbf{q}
U^e(\textbf{p}-\textbf{q},\textbf{q})U^e(\textbf{q}-\textbf{p},\textbf{p})
P(\textbf{q},0;\omega,0)\right].
\end{equation}
A spherically symmetric solution satisfies
\begin{equation}\label{Pdef}
P(p,0;\omega,0)=G^2(p, \omega)\left[1+\frac{u}{4} \epsilon_B^2 e^{-
p^2 / 2} \int^{\infty}_0 \ud q \ q e^{- q^2 / 2} (I_0 (q p)-1 )
P(q,0;\omega,0)\right].
\end{equation}
Let us introduce a new function
\begin{equation}
H(p,\omega)\equiv P(p,0;\omega,0)G^{-2}(p, \omega);
\end{equation}
then Eq.\;(\ref{Pdef}) can be rewritten as
\begin{equation}
H(p,\omega)=1+\frac{u}{4} \epsilon_B^2 e^{- p^2 / 2}
\int^{\infty}_0 \ud q \ q e^{- q^2 / 2} (I_0 (q p)-1 )G^2(q,
\omega) H(q,\omega),
\end{equation}
or explicitly
\begin{equation}\label{intgequ}
H(p,\omega)=1+\frac{u}{4} \epsilon_B^2 e^{- p^2 / 2}
\int^{\infty}_0 \ud q \ q e^{- q^2 / 2} \frac{I_0 (q p)-1
}{[\omega - \omega_q - \Sigma_u (q, \omega) ]^2} H(q,\omega).
\end{equation}
Notice that Eq.\;(\ref{intgequ}) has the same form as
Eq.\;(\ref{selfderiv}) but with $H(p,\omega)$ instead of
$\partial_u\Sigma_u (p, \omega)$, which is known to diverge
$\partial_u\Sigma_u (p, 0)\rightarrow \infty$ when $u \rightarrow
u_c$. In the next part we will demonstrate that $H(p,0)$ also
diverges, $H(p,0)\rightarrow \infty$ when $u \rightarrow u_c$.
\end{widetext} We are looking for a solution in the form
\begin{equation}\label{solform}
H(p,\omega) = 1+e^{- p^2 / 2}\sum_{n=1}^{\infty}h_n(\omega)p^{2n},
\end{equation}
Substitution of Eq.\;(\ref{solform}) into Eq.\;(\ref{intgequ})
yields an expression, which looks similar to the equation previously
obtained (see Eq.\;(\ref{matreq})),
\begin{equation}\label{matreq2}
h_n(\omega) = \frac{uK_{n}}{4(2^n n!)^2}+\frac{u}{4(2^n
n!)^2}\sum_{k=1}^{\infty}F_{n+k}h_k(\omega),
\end{equation}
where the function $F_{n}$ was defined earlier by Eq.\;(\ref{Fndef})
and
\begin{equation}
K_n \equiv \epsilon_B^2 \int^{\infty}_0 \ud q \  \frac{q^{2n+1}
e^{- q^2/2} }{[\omega - \omega_{q} - \Sigma_u (q, \omega)]^2}.
\end{equation}
Notice that
\begin{equation}
K_n =\sum_{k=0}^{\infty}\frac{1}{2^k k!}F_{k+n}
\end{equation}
and
\begin{equation}
F_n =\sum_{k=0}^{\infty}\frac{(-1)^k}{2^k k!}K_{k+n}.
\end{equation}
 Equivalently
\begin{equation}\label{matreq3}
\sum_{k=1}^{\infty}B_{n,k}2^k k!h_k(\omega)=\frac{uK_{n}}{2^{n+2}
n!},
\end{equation}
where $B_k,n$ was defined in Eq.\;(\ref{matr}). The solution is
found by computing the inverse matrix to Eq.\;\eqref{matreq3} and
has the form
\begin{equation}\label{matrsol2}
h_n(\omega)=\frac{2^{-n}u}{4
n!}\sum_{k=1}^{\infty}B^{-1}_{n,k}\frac{K_{k}}{2^{k} k!}.
\end{equation}
\begin{widetext}
Therefore,
\begin{equation}
P(p,0;\omega,0)=G^2(p, \omega)\left( 1+e^{- p^2
/2}\sum_{n=1}^{\infty}p^{2n}\frac{2^{-n}u}{4
n!}\sum_{k=1}^{\infty}B^{-1}_{n,k}\frac{K_{k}}{2^{k} k!} \right)
\end{equation}
and
\begin{equation}
\int_{0}^{\infty} P(p,0;\omega,0)p \; \ud p=\int_{0}^{\infty}
G^2(p, \omega)p \; \ud p+\sum_{n=1}^{\infty}\int_{0}^{\infty} e^{-
p^2 /2}p^{2n+1}G^2(p, \omega)\ud p \frac{2^{-n}u}{4
n!}\sum_{k=1}^{\infty}B^{-1}_{n,k}\frac{K_{k}}{2^{k} k!},
\end{equation}
if the integral is convergent. Otherwise, it has to be
regularized, which we won't consider here. This leads to
\begin{equation}
\int_{0}^{\infty} P(p,0;\omega,0)p \; \ud
p=\sum_{n=1}^{\infty}\frac{K_{n}}{2^{n}
n!}+\frac{u}{4}\sum_{n,k=1}^{\infty}\frac{K_{n}}{2^{n}
n!}B^{-1}_{n,k}\frac{K_{k}}{2^{k} k!},
\end{equation}
which can be further simplified by means of some algebraic
transformations,
\begin{equation}
\int_{0}^{\infty} P(p,0;\omega,0)p \; \ud
p=F_0+\sum_{n,k=1}^{\infty}\left(\frac{4}{u}(B^{-1}_{n,k}-\delta_{n,k})+2B^{-1}_{n,k}\frac{F_{k}}{2^{k}
k!}+\frac{u}{4}\frac{F_{n}}{2^{n} n!}B^{-1}_{n,k}\frac{F_{k}}{2^{k}
k!}\right).
\end{equation}
\end{widetext} Despite the simplifications, the above expression is
difficult to evaluate analytically, as well as numerically.
However, since most of the terms there involve the inverse matrix,
it is reasonable to suppose that if $\omega=0$ it diverges with $u
\rightarrow u_c$ as
\begin{equation}
\int_{0}^{\infty} P(p,0;0,0)p \; \ud p \sim \det[B(u)]^{-1}.
\end{equation}
On the other hand the susceptibility is given by
\begin{equation}
\chi(\epsilon)=-\frac{iA}{(2\pi)^2}\int_{-\infty}^{\infty} \ud
\omega \int_{0}^{\infty} P(p,0;\omega,\epsilon)p \; \ud p.
\end{equation}
Thus, considering $\epsilon=0$, we see that the integrand is
divergent at $\omega=0$ with $u \rightarrow u_c$, which is
definitely not enough to infer the divergence of the integral
itself, but can be considered as an indication to such
possibility.


\begin{thebibliography}{99}

\bibitem{green} A.\ G.\ Green, \textit{Phys. Rev.} B \textbf{57}, R9373 (1998).


\bibitem{macdgirvin} J.\ Sinova, A.\ H.\ MacDonald and S.\ M.\ Girvin, \textit{Phys. Rev.} B \textbf{62}, 13579 (2000).

\bibitem{chalkerlett} S.\ Rapsch, J.\ T.\ Chalker and D.\ K.\ K.\ Lee, \textit{Phys. Rev. Lett.} \textbf{88}, 036801 (2002); D.\ K.\ K.\ Lee, S.\ Rapsch and J.\ T.\ Chalker, \textit{Phys. Rev.} B \textbf{67}, 195322 (2003).

\bibitem{tomonaga} J.\ Tomonaga, \textit{Progr.\ Theor.\ Phys.}
\textbf{5}, 544-569 (1950).

\bibitem{intro2} A.\ Luther, \textit{Phys. Rev. B} \textbf{19}, 320 (1979).

\bibitem{intro3} F.\ D.\ M.\ Haldane, \textit{Helv. Phys. Acta} \textbf{65}, 152 (1992).

\bibitem{intro4} A.\ H.\ Castro Neto, E.\ Fradkin, \textit{Phys. Rev. B} \textbf{49}, 10877 (1994).

\bibitem{intro5} A.\ Houghton and B.\ Marston, \textit{Phys. Rev.} B \textbf{48}, 7790 (1993).

\bibitem{intro6} H.\ J.\ Kwon, A.\ Houghton and B.\ Marston, \textit{Phys. Rev.} B \textbf{52}, 8002 (1995).

\bibitem{D2} H.\ Westfahl Jr., A.\ H.\ Castro Neto, and A.\ O.\ Caldeira, \textit{Phys. Rev.} B \textbf{55}, R7347 (1997).

\bibitem{doretto} R.\ L.\ Doretto, A.\ O.\ Caldeira and S.\ M.\ Girvin, \textit{Phys. Rev.} B \textbf{71}, 045339 (2005).

\bibitem{Lazarides}O.\ Tieleman, A.\ Lazarides, D.\ Makogon, and C.\ Morais Smith, in preparation.

\bibitem{kallin} C.\ Kallin and B.\ I.\ Halperin, \textit{Phys. Rev.} B {\bf 30}, 5655 (1984).

\bibitem{Girvin}S.\ M.\ Girvin, A.\ H.\ MacDonald, and P.\ M.\ Platzman, \textit{Phys. Rev.} B \textbf{33}, 2481 (1986).

\bibitem{fogler} M.\ M.\ Fogler and B.\ I.\ Shklovskii, \textit{Phys. Rev.} B \textbf{52}, 17 366 (1995).

\bibitem{Avgin}I.\ Avgin, D.\ L.\ Huber, and W.\ Y.\ Ching, Phys. Rev. B {\bf 48}, 16109 (1993).

\bibitem{Shender}E.\ F.\ Shender, J.\ Phys., C {\bf 11}, L423 (1978).

\bibitem{MacDonald}A.\ H.\ MacDonald, P.\ M.\ Platzman, and G.\
S.\ Boebinger, \textit{Phys. Rev. Lett.} \textbf{65}, 775 (1990).

\bibitem{murthy} G.\ Murthy, Phys. Rev. B \textbf{64}, 241309 (2001).

\bibitem{doniach} H.\ Doniach, E.\ H.\ Sondheimer, \textit{Green's Functions for
Solide State Physicists}, (W.\ A.\ Benjamin, Inc., 1974).

\bibitem{bruus} H.\ Bruus, K.\ Flensberg, \textit{Many-Body Quantum Theory in Condensed-Matter Physics}, (Oxford University Press, 2006).

\bibitem{fertigmurthy} H.\ A.\ Fertig, G.\ Murthy, \textit{Phys. Rev. Lett.} \textbf{95}, 156802 (2005).




\bibitem{gradshteyn} I.\ S.\ Gradshteyn and I.\ M.\ Ryzhik, \textit{Table of Integrals, Series and Products}, (Academic Press, San Diego) 1994.






\end{thebibliography}
\end{document}